\documentclass[twocolumn,showpacs,aps,prb]{revtex4}

\usepackage{dcolumn}
\usepackage{epsfig}
\def\gsim{\; $\raise0.3ex\hbox{$>$}\llap{\lower0.8ex\hbox{$\sim$}}$\;}
\def\lsim{\; $\raise0.3ex\hbox{$<$}\llap{\lower0.8ex\hbox{$\sim$}}$\;}

\begin{document}
\title{Fractional $S^z$ excitation and its bound state around the 1/3 plateau of the S=1/2 Ising-like  zigzag XXZ chain.}
\author{Kouichi Okunishi and  Takashi Tonegawa${}^1$}
\affiliation{
Department of Physics, Faculty of Science, Niigata University, Igarashi 2, 950-2181, Japan\\
${}^1$Department of Mechanical Engineering, Fukui University of Technology, 6-1 Gakuen 3-Chome, Fukui-Shi 910-8505, Japan.
}
\date{\today}

\begin{abstract}
We present the microscopic view for the excitations around the 1/3 plateau state of the Ising-like zigzag XXZ chain.
We analyze the low-energy excitations around the plateau with  the degenerating perturbation theory from the Ising limit, combined with the Bethe-form wave function.
We then find that the domain-wall particles carrying $S^z=\pm 1/3$ and its bound state of $S^z=\pm 2/3$ describe well the low-energy excitations around the 1/3 plateau state.
The formation of the bound state of the domain-walls clearly provides the microscopic mechanism of  the cusp singularities and the even-odd behavior in the magnetization curve.
\end{abstract}


\maketitle

\section{introduction}

The microscopic view of the low-energy excitation in frustrated quantum spin systems has been on a field of the active researches, since the structure of the low-energy excitation directly reflects on the experimentally observable quantities such as magnetization curve.
For instance, various interesting behaviors of the magnetization curve are observed experimentally for such a system as SrCu(BO$_3$)$_2$\cite{ss} and the exotic nature of the low-energy excitation is actually reported behind the characteristic magnetization curve\cite{mt}.
However, in general, it is not easy to reveal the role of the frustration in the low-dimensional quantum spin systems theoretically, where the quantum fluctuation and frustration affects the low-energy excitation cooperatively.

Among various frustrating spin systems, the $S=1/2$ zigzag XXZ chain is one of the most fundamental models having the frustrating interaction without loss of the translational invariance\cite{mg,halzig,tone1,nomura,allen,sen,whiteaffleck}, and is actually synthesized as SrCuO$_2$\cite{matsuda}, Cu(ampy)Br$_2$\cite{kikuchi}, F$_2$PIMNH\cite{hosokoshi} and (N$_2$H$_5$)CuCl$_3$\cite{haginaru}. 
The Hamiltonian of the zigzag chain has a very simple structure, and has been playing a crucial role for the purpose of understanding the frustration effect, since it captures a variety of interesting behaviors induced by the frustration.
In fact, the zigzag chain in a magnetic field has been studied actively.\cite{tone2,schmidt,cabra,cusp,maeshima,mobius}
Very recently, we have presented the exotic magnetic phase diagram of the zigzag chain including the strongly frustrated region;\cite{OT} we have found the magnetization plateau at 1/3 of the full moment accompanying the spontaneous breaking of the translational symmetry, the cusp singularities, and the interesting even-odd effect in the magnetization curve.

Although the above exotic properties of the magnetization curve have been illustrated by the extensive numerical calculations, the microscopic mechanism for them around the 1/3 plateau has not been investigated systematically.
In order to address such microscopic views, we focus on the two key features of the zigzag chain: 
the first one is  that the $\uparrow\uparrow\downarrow$ spin structure is realized at the 1/3 plateau state, which is extended up to the Ising limit of the zigzag XXZ chain\cite{OT,morihori}. 
Another one is that the Hamiltonian of the zigzag chain interpolates between the single Heisenberg chain and the double Heisenberg chain continuously  without loss of the translational invariance.
How can we connect these two features with the exotic magnetization curve of the isotropic zigzag chain?
A more thorough understanding of the problem is clearly desirable, which will provide an essential insight for the effect of the frustration in the zigzag-structured system.

In this paper, we study the microscopic picture of the low-lying excitation  around 1/3 plateau for the zigzag XXZ chain.
We first classify the excitation on the 1/3 plateau in the Ising limit and then take account of the quantum effect with the degenerating perturbation theory.
We find that the low-energy excitation can be represented as the combination of the domain-wall(DW) type excitations carrying $S^z=\pm 1/3$ and $\pm 2/3$, where the DW particle of $S^z=\pm 2/3$ can be regarded as the bound state of the two $S^z=\pm 1/3$ particles. 
We further analyze the two-body problem of the $S^z=\pm 1/3$ DWs invoking the Bethe type wave function\cite{kiwata}, and then clarify that the formation of the DW bound state becomes important as the double-chain nature of the system  becomes dominant.
Calculating the magnetization curve with the density-matrix renormalization group(DMRG) method\cite{dmrg}, we further delineate that the DW particle picture explains well the low-energy excitation and  the magnetization curve around the 1/3 plateau state.
In addition to these, we report that the DW defect can be inserted even in the 1/3 plateau state for a finite size system with the open boundary condition(OBC).

This paper is organized as follows:
In the next section, we introduce the zigzag XXZ chain  briefly.
In Sec.III, we describe the 1/3 plateau state of the zigzag chain in the Ising limit. The basic structure of the DW excitation around the plateau is explained here.
In Sec.IV, we take account of the XY term with the degenerating perturbation theory. In order to solve the eigenvalue problem for the two DW problem, we invoke the Bethe type wavefunction and clarify the bound-state formation condition.
In Sec. V the DMRG result is presented and the relation with the analytical results are discussed. 
In Sec. VI, we discuss the effect of the OBC on the 1/3 plateau state.
In Sec. VII, the conclusions are summarized.
We also make a comment on the recently synthesized zigzag compound (N$_2$H$_5$)CuCl$_3$\cite{haginaru}. 

\section{model}

The model we consider here is the Ising-like XXZ chain with the nearest neighbor(NN) coupling $J_1$ and the next-nearest neighbor(NNN) one $J_2$ in a magnetic field $H$. The Hamiltonian of the model is given by
\begin{eqnarray}
{\cal H}=  {\cal H}_0 + {\cal H}_{\rm zeeman},
\label{zigzag}
\end{eqnarray}
where ${\cal H}_0$ is the Hamiltonian of the system part and ${\cal H}_{\rm zeeman}\equiv  -H \sum_{i=1}^{3N} S^z_i$ is the Zeeman interaction term.
The Hamiltonian of the system part consists of the NN and NNN  interaction terms: ${\cal H}_0\equiv {\cal H}_{\rm NN}+{\cal H}_{\rm NNN}$ with
\begin{eqnarray}
 {\cal H}_{\rm NN}&=& J_1 \sum_{i=1}^{3N} h_{i,i+1} , \quad  {\cal H}_{\rm NNN}=  J_2\sum_{i=1}^{3N} h_{i,i+2}, \\
 h_{i,j} &=& \varepsilon (S^x_iS^x_{j}+S^y_iS^y_{j} )+S^z_iS^z_{j}
\end{eqnarray}
where $\vec{S}$ is the $S=1/2$ spin operator, and  $\varepsilon$ denotes the anisotropy of the XY term. 
In the paper, we consider the system of $3N$ sites with $N=$even.
We also introduce the notation $\alpha\equiv J_2/J_1$ for simplicity.

\begin{figure}[hb]
\epsfig{file=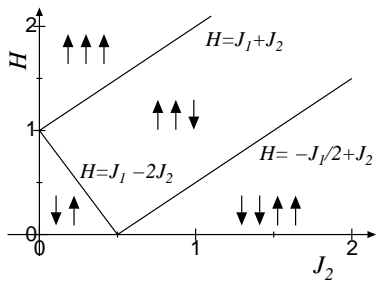,width=5cm}
\caption{The magnetic phase diagram of the zigzag Ising chain obtained in Ref.[\cite{morihori}]. In the figure, $J_1$ is normalized to be unity.} \label{ilimit}
\end{figure}

In the Ising limit($\varepsilon=0$), the phase diagram at zero temperature was obtained by Morita and Horiguchi about 30 years ago.\cite{morihori} The 1/3 plateau appears in the region denoted as $\uparrow\uparrow\downarrow$ in the Fig.\ref{ilimit}.
The regions denoted as $\uparrow\downarrow$ and $\uparrow\uparrow\downarrow\downarrow$ mean the N\'eel and double-N\'eel phases respectively, both of which have zero magnetization.
An interesting point in the phase diagram is that the phase boundary of the 1/3 plateau merges the boundary of the N\'eel and double-N\'eel transition point at $\alpha=1/2$.
This implies that the system is highly degenerating at $\alpha=1/2$.

For the isotropic case($\varepsilon=1$), the phase diagram including the strongly frustrated region is recently obtained by the authors\cite{OT};
 the 1/3 plateau survives for $0.56 <\alpha <1.25 $ in spite of the strong quantum fluctuation.
In addition, the cusp singularities show interesting behaviors around the 1/3 plateau.
Moreover, the even-odd behavior of the magnetization curve appears in a large $J_2$ region.
When analyzing these interesting  magnetization curve, a key feature of the zigzag chain  is that the frustration effect can be casted as the interplay between the single chain and double chain natures.
For example, the two component Tomonaga-Luttinger(TL) liquid is realized below the low-field cusp and/or above the high-filed cusp, reflecting the two chain nature of the system, while the middle-field blanch consists of the one component TL liquid.\cite{cusp}

However the microscopic mechanism of these behaviors has not been illustrated yet, since the incommensurate nature of the system due to the frustration is difficult to treat theoretically.
In what follows, we address the problem starting from the Ising limit of the zigzag XXZ chain.

\section{Ising limit}

In this section , we investigate the nature of the excitations around the 1/3 plateau state of the zigzag XXZ chain in the Ising limit.
Although the argument in the Ising limit is quite simple, we can illustrate the basic aspects of the excitations around the plateau,  providing a good starting point for taking account of the quantum effect with the perturbative argument of the XY term.

\subsection{1/3 plateau state}

We first remark the 1/3 plateau state in the Ising limit briefly.
The ground state phase diagram of the zigzag Ising chain was obtained by Morita and Horiguchi.\cite{morihori}
As soon as $J_2$ is introduced, the 1/3 plateau appears in the region denoted as $\uparrow\uparrow\downarrow$ in Fig.\ref{ilimit}.
Here we should recall that the Hamiltonian is translational invariant;
at the 1/3 plateau state, one of the following three configuration is chosen, accompanying the spontaneous symmetry breaking of the translation:
\begin{eqnarray*}
|{\rm N\acute{e}el}_1\rangle&=&\cdots \uparrow\downarrow\uparrow\uparrow\downarrow\uparrow\uparrow\downarrow\uparrow\uparrow\downarrow\uparrow \cdots \\
|{\rm N\acute{e}el}_2\rangle&=&\cdots \uparrow\uparrow\downarrow\uparrow\uparrow\downarrow\uparrow\uparrow\downarrow\uparrow\uparrow\downarrow \cdots \\
|{\rm N\acute{e}el}_3\rangle&=&\cdots \downarrow\uparrow\uparrow\downarrow\uparrow\uparrow\downarrow\uparrow\uparrow\downarrow\uparrow\uparrow \cdots .
\end{eqnarray*}

For the system of $3N$ sites with the periodic boundary condition(PBC), the energy of the above N\'eel states is 
\[ E_g=-\frac{N}{4}(J_1+J_2), \]
where we have neglected the energy of the Zeeman term. The total magnetization of the state is trivially $M=N$(or the magnetization per site is $m= M/3N  =1/3$).

\subsection{excitation around the plateau}

In order to analyze the low-energy excitations around the 1/3 plateau state, it is instructive to recall the excitation of the XXZ chain(without NNN term) at zero magnetic field, which is described by the DW separating  the two-fold degenerating spin arrays of the N\'eel-ordered ground state.\cite{ishimurashiba}
For the present case, the excitation on the plateau state is described by the DW that is made from the combinations of the three N\'eel-ordered spin aligns. 
Since the plateau state is three fold degenerating, there are 6(=$_3$P$_2$)  possible types of the DW, which are summarized in TABLE I.
As will be seen soon, the DW can be regarded as a particle carrying fractional value of $S^z$. 
The energy of the DW particle measured from the 1/3 plateau state is also listed in TABLE I.

\begin{table}[bt]
\begin{ruledtabular}
\begin{tabular}{ccrc}
  type & spin array   &  $S^z$       &energy   \\ \hline
 1-2 & $\cdots\uparrow\downarrow\uparrow\uparrow\uparrow\downarrow\cdots$ & 1/3   &$ \frac{1}{3}(J_1+J_2)$\\ 
 3-1 & $\cdots\downarrow\uparrow\uparrow\uparrow\downarrow\uparrow\cdots$ & 1/3   &$  \frac{1}{3}(J_1+J_2)$\\
 3-2 & $\cdots\downarrow\uparrow\uparrow\uparrow\uparrow\downarrow\cdots$ & 2/3   &$  \frac{2}{3}(J_1+J_2)$\\
 1-3 & $\cdots\uparrow\downarrow\uparrow\downarrow\uparrow\uparrow\cdots$ & -1/3  &$  \frac{1}{3}(-J_1+2J_2)$\\
 2-1 & $\cdots\uparrow\uparrow\downarrow\uparrow\downarrow\uparrow\cdots$ & -1/3  &$  \frac{1}{3}(-J_1+2J_2)$\\
 2-3 & $\cdots\uparrow\uparrow\downarrow\downarrow\uparrow\uparrow\cdots$ & -2/3  &$  \frac{1}{3}(J_1-2J_2)$
\end{tabular}
\end{ruledtabular}
\caption{The low energy excitation on the 1/3 plateau state, which is described by the domain wall made from three type of the N\'eel ordered spin aligns. These domain-walls can be regarded as a quasi-particles carrying fractional $S^z$.
The numbers noted in the ``type'' column indicate the three types of  N\'eel ordered spin aligns.}
\end{table}

In order to see the origin of the fractional value of $S^z$, we start with the wave function for $M=N+1$ sector that is created by operating $S^+$ for a site in the 1/3 plateau state:
\begin{eqnarray*}
| \psi^+_{00}\rangle=\cdots \downarrow\uparrow\uparrow\downarrow\uparrow\uparrow\!\!\!\>\dot{\uparrow}\!\!\!\>\uparrow\uparrow\downarrow\uparrow\uparrow\downarrow\uparrow \cdots \label{psi00}
\end{eqnarray*}
where ``dot'' is assigned for the inverted spin.
The energy of the $|\psi^+_{00}\rangle$ state is  
\[ E^+=E_g+(J_1+J_2).\]
The $|\psi^+_{00}\rangle$  state is not represented as a simple DW.
By exchanging the nearest neighboring $\uparrow$ and  $\downarrow$ spins,  we can decompose the five up spin cluster into combinations of the fundamental DWs and thus generate the low-energy excited state for the  $M=N+1$ sector systematically:
\begin{eqnarray}
| \psi^+_{01}\rangle&=&\cdots \downarrow\uparrow\uparrow\downarrow\!\!\!\>\underline{\uparrow\uparrow\uparrow\uparrow}\!\!\!\>\downarrow\!\!\!\>\underline{\uparrow\uparrow\uparrow}\!\!\!\>\downarrow\uparrow\uparrow\downarrow\uparrow \cdots \nonumber \\
| \psi^+_{02}\rangle&=&\cdots \downarrow\uparrow\uparrow\downarrow\!\!\!\>\underline{\uparrow\uparrow\uparrow\uparrow}\!\!\!\>\downarrow\uparrow\uparrow\downarrow\!\!\!\>\underline{\uparrow\uparrow\uparrow}\!\!\!\>\downarrow\uparrow \cdots \nonumber \\
| \psi^+_{12}\rangle&=&\cdots \downarrow\!\!\!\>\underline{\uparrow\uparrow\uparrow}\!\!\!\>\downarrow\!\!\!\>\underline{\uparrow\uparrow\uparrow}\!\!\!\>\downarrow\uparrow\uparrow\downarrow\!\!\!\>\underline{\uparrow\uparrow\uparrow}\!\!\!\>\downarrow\uparrow \cdots \nonumber  \\
&\vdots& \nonumber \\
| \psi^+_{lm}\rangle&=&\cdots \uparrow\downarrow\!\!\!\>\underline{\uparrow
}\!\overbrace{\underline{\uparrow\uparrow}\!\!\!\>\downarrow\cdots
\uparrow\uparrow\downarrow}^{3l\; {\rm sites}}\underline{\uparrow\uparrow\uparrow }
\overbrace{\downarrow\uparrow\uparrow \cdots 
\downarrow\!\!\!\>\underline{\uparrow\uparrow}}^{3m\;\rm{sites}}\!\underline{\uparrow }\!\!\!\>\downarrow\uparrow \cdots  \label{updwarray}
\end{eqnarray}
where we assign underlines to the $\uparrow\uparrow\uparrow$ or $\uparrow\uparrow\uparrow\uparrow$ clusters that are useful signs to find the DWs.
For all configurations above, we can easily confirm that the energy is the same as $E^+=E_g+(J_1+J_2)$.
These degenerating states are connected by of XY term, which is essential for the degenerating perturbation treatment in the next subsection.

By comparing the spin configurations in eq. (\ref{updwarray})  with those in TABLE I,  we can identify the $\uparrow\uparrow\uparrow$ cluster as the DW carrying $S^z=1/3$; 
Since all of the $\uparrow\uparrow\uparrow$ clusters in $| \psi^+_{lm}\rangle$ state are equivalent, the three DWs are sharing the magnetization $S^z=1$ equally.
Thus each DW is represented as a particle of $S^z=1/3$ and $\Delta E= (J_1+J_2)/3$.\cite{vaccumene}
In addition to this, the configuration of $\uparrow\uparrow\uparrow\uparrow$ can be regarded as a bound state of the two $S^z=1/3$ DWs, which has $S^z=2/3$ and  $\Delta E=2(J_1+J_2)/3$. 
As is described in the next section, this bound state becomes important for $J_2\gg J_1$.

Also for $M=N-1$ sector, we can make almost the parallel argument to the $S^z=1/3 $ DW case.
Inverting an up spin in the 1/3 plateau state, we have a state:
\begin{eqnarray*}
| \psi^-_{\bar{0}0}\rangle=\cdots \downarrow\uparrow\uparrow\downarrow\uparrow\uparrow\downarrow\uparrow\uparrow\downarrow\!\!\!\>\dot{\downarrow}\!\!\!\>\uparrow\downarrow\uparrow\uparrow\downarrow\uparrow\uparrow\downarrow\uparrow \cdots ,
\end{eqnarray*}
whose energy is $E^-=E_g$.
Although this value of the energy is the same as that of the 1/3 plateau state, the total energy including the Zeeman terms for the 1/3 plateau state is lower than that for $| \psi^-_{\bar{0}0}\rangle$ in the certain range of the magnetic field. 

We also generate the low-energy  states for the $M=N-1$ sector, by exchanging the nearest neighboring $\uparrow$ and $\downarrow$ spins.
However, we note that there is a bit different point from the $M=N+1$ case.
Since the state $| \psi^-_{\bar{0}0}\rangle$ contains two kind of the spin clusters, $\downarrow\downarrow$ and $\downarrow\uparrow\downarrow$, 
 we introduce the slightly modified labeling for the position of the  $\downarrow\downarrow$ cluster; we assign ``bar'' to the label of the $\downarrow\downarrow$ configuration.
\begin{eqnarray*}
| \psi^-_{\bar{0}1}\rangle&=&
\cdots \downarrow\uparrow\uparrow\downarrow\uparrow\uparrow\downarrow\uparrow\uparrow\!\!\!\>\underline{\downarrow\downarrow}\!\!\!\>\uparrow\uparrow\!\!\!\>\underline{\downarrow\uparrow\downarrow}\!\!\!\>\uparrow\uparrow\downarrow\uparrow \cdots \\
| \psi^-_{01}\rangle&=&
\cdots \downarrow\uparrow\uparrow\downarrow\uparrow\uparrow\!\!\!\>
\underline{\downarrow\uparrow\downarrow\uparrow\downarrow}\!\!\!\>
\uparrow\uparrow\!\!\!\>\underline{\downarrow\uparrow\downarrow}\!\!\!\>\uparrow\uparrow\downarrow\uparrow \cdots \\
| \psi^-_{11}\rangle&=&
\cdots \downarrow\uparrow\uparrow\!\!\!\>\underline{\downarrow\uparrow\downarrow}\!\!\!\>\uparrow\uparrow\!\!\!\>
\underline{\downarrow\uparrow\downarrow}\!\!\!\>
\uparrow\uparrow\!\!\!\>\underline{\downarrow\uparrow\downarrow}\!\!\!\>\uparrow\uparrow\downarrow\uparrow \cdots \\
| \psi^-_{21}\rangle&=&
\cdots \underline{\downarrow\uparrow\downarrow}\!\!\!\>
\uparrow\uparrow\downarrow\uparrow\uparrow
\!\!\!\>\underline{\downarrow\uparrow\downarrow}\!\!\!\>
\uparrow\uparrow
\!\!\!\>\underline{\downarrow\uparrow\downarrow}\!\!\!\>\uparrow\uparrow\downarrow\uparrow \cdots \\
&\vdots& \\
| \psi^-_{lm}\rangle&=&
\cdots \underline{\downarrow\uparrow }
\!\overbrace{\underline{\downarrow} \!\!\!\>\uparrow\uparrow
\cdots
\downarrow\uparrow\uparrow}^{3l\; {\rm sites}}
\!\underline{\downarrow\uparrow\downarrow}\!
\overbrace{\uparrow\uparrow\downarrow
\cdots
\uparrow\uparrow\!\!\!\>\underline{\downarrow}}^{3m\; {\rm sites}}
\!\underline{\uparrow\downarrow}\!\!\!\>\uparrow \cdots 
\end{eqnarray*}
where underlines are assigned for the $\downarrow\uparrow\downarrow$ and  $\downarrow\downarrow$ clusters.
Clearly the $\downarrow\uparrow\downarrow$ cluster can be regarded as a DW-particle carrying $S^z=-1/3$, corresponding to the type 1-3 or 2-1 DWs in TABLE 1.
Moreover, the energy for the states consisting of the three $\downarrow\uparrow\downarrow$ DWs is easily calculated as $E=E_g-J_1+2J_2$.
Thus the energy of the $S^z=-1/3$ DW particle reads $\Delta E=(-J_1+2J_2)/3$.

We further consider the $\downarrow\downarrow$ cluster, which is depicted as the 2-3 type DW in TABLE 1. 
Since the  $\downarrow\downarrow$ cluster is decomposed into the two $\downarrow\uparrow\downarrow$ clusters,  the $\downarrow\downarrow$ cluster can be regarded as the bound state of the two $S^z=-1/3$ DWs.
Thus we find that the $\downarrow\downarrow$ cluster is the DW bound state carrying $S^z=-2/3$.
In addition, we note that the total energy for the state including the $S^z=-2/3$ DW(for example $|\psi^-_{\bar{0}1}\rangle$) is $E=E_g$.
Therefore the energy of the  $S^z=-2/3$ DW bound state is readily identified to be $\Delta E=(J_1-2J_2)/3$.

An essential point on the 1/3 plateau state is that the low-energy excitations of the $M=N\pm 1$ sector are described by the combinations of the DWs listed in TABLE 1.  We next take account of  the quantum effect originating from the XY term on the basis of the DW picture.

\section{perturbation by the $XY$-term}

In this section, we consider the the quantum effect on the DW excitations illustrated in the Ising limit, by using the degenerating perturbation theory with respect to the XY term. We assume $ \varepsilon \ll 1$, for which the perturbation treatment is justified.
Here we should note that the correction to the energy of the 1/3 plateau state starts with  2nd order of $\varepsilon$:
\begin{equation}
E_{g}/N = -\frac{1}{4}(J_1+J_2) - \frac{\varepsilon^2}{2}\left(\frac{J_1^2}{J_2}+\frac{J_2^2}{J_1}\right).
\end{equation}
Thus we can set the origin of the energy as $E_g= -\frac{N}{4}(J_1+J_2)$ in the following arguments of the first order degenerating perturbation for the DW excitations.

\subsection{basic views for the DW}

We  analyze the one-body problem of $S^z=1/3$ DW in the infinite length chain up to the first order of $\varepsilon$. 
Although the higher order terms may be required for  a quantitative analysis of the problem, we can capture the intrinsic property of the DW excitation within the first order theory.

We label the position of the $S^z=1/3$ DW as $x$ which is defined as the site of the center of three up spins.\cite{position}
Then the matrix element of the ${\cal H}_0$ is obtained as
\begin{eqnarray}
{\cal H}_0 | x \rangle &=&\frac{J_1+J_2}{3} |x\rangle + \frac{\varepsilon J_1 }{2}\left( |x-3 \rangle+|x+3\rangle \right)\nonumber \\
& &+ {\rm higher\; energy\; terms} 
\end{eqnarray}
where the higher energy terms include the configurations that yield the energy rise of order $J_1$ or $J_2$.
Here an important point is that the NNN term always generates higher energy terms.
Thus the NNN term can not contribute to the low-energy dynamics of the single DW within the first order of $\varepsilon$.
By neglecting the higher energy terms, we can easily obtain the dispersion curve  of the single DW: 
\begin{equation}
\omega(k)= \frac{1}{3}(J_1+J_2)+ \varepsilon J_1\cos( 3 k) .
\end{equation}

As was seen above, if the NNN term is operated on the single DW state, it generates only higher energy configurations.
Then a natural question arises: what is the role of the NNN term?
In order to see it, we examine the effect of the NNN term on the DW bound state of $S^z=2/3$.
We define the state for the DW bound state at the position $z$ by
\begin{equation}
|z\rangle\equiv \cdots \downarrow\uparrow\uparrow\downarrow\uparrow\!\!\!\>\mathop{\uparrow}_{z}\!\!\!\>\uparrow\uparrow\downarrow\uparrow\uparrow\downarrow \cdots .
\end{equation}
Then the matrix element of ${\cal H}_{\rm NNN}$ for the DW bound state is
\begin{eqnarray}
{\cal H}_{\rm NNN} | z \rangle &=&\frac{2}{3}(J_1+J_2) |z\rangle + \frac{\varepsilon J_2 }{2}\left( |z-3 \rangle+|z+3\rangle \right)\nonumber \\
& &+ {\rm higher\; energy\; terms}  \label{updwhop}
\end{eqnarray}
where the higher energy configurations are generated by the NNN term operated on the spins away from the DW part.
Thus we see that the DW bound state can move by using the NNN term without the extra energy cost.
This is a crucial point on the NNN term. 
When $J_1\gg J_2$, the low-energy excitation on the plateau is basically described by the single DW excitation.
On the other hand, when $J_1 \ll J_2$, we can expect that the role of the DW bound state becomes essential in the low-energy excitation.
The switching mechanism between the single DW excitation and the DW bound state is  clearly associated with the crossover between the single chain nature and double chain nature of the zigzag chain. 
In the next subsection, we analyze the two DWs and its bound state problem systematically.

\subsection{two body problem and bound state}

In order to understand the formation mechanism of the DW bound state,
we consider the two body problem of the DWs in the infinite chain.\cite{kiwata}
Introducing the notation for the free two DW state,
\[
|x,y\rangle= \cdots \uparrow\downarrow\uparrow \!\!\!\>\mathop{\uparrow}_{x}\!\!\!\> \uparrow\downarrow\uparrow \cdots \uparrow\downarrow\uparrow\!\!\!\>\mathop{\uparrow}_{y}\!\!\!\>\uparrow\downarrow\uparrow  \cdots
\]
we write the wave function as
\begin{equation}
|\psi \rangle = \sum_{x<y} f_{x,y}|x,y\rangle +\sum_z g_z|z\rangle
\end{equation}
where the summentions about $x,y,z$ are taken over the all possible positions of the DWs.
Then we obtain the real-space eigenvalue equation for the two DW scattering problem:
\begin{widetext}
\begin{eqnarray}
& & E^{(2)} f_{x,x+4}
=
\frac{2}{3}(J_1+J_2)f_{x,x+4}
+\frac{\varepsilon J_1}{2}\left[ f_{x-3,x+4}+f_{x,x+7} 
+ g_x+ g_{x+3} \right]\;,  \label{ev1}\\ 
& & E^{(2)} g_x = 
\frac{2}{3}(J_1+J_2)g_x
+\frac{\varepsilon J_2}{2}\left[ g_{x+3}+g_{x-3} \right]
+\frac{\varepsilon J_1}{2}\left[ f_{x-3,x+1}+f_{x,x+4} \label{ev2}\right]\;.
\end{eqnarray}
In order to deal with eqs. (\ref{ev1}) and (\ref{ev2}), we assume the Bethe type wave function for the two DW problem:
\begin{equation}
f_{x,y}=\sum_{k,k'} A(k,k')e^{ikx+ik'y}+B(k,k')e^{iky+ik'x}. \label{bethe2}
\end{equation}
We also write the wave function for the bound state as
\begin{equation}
g_x=\sum_{k,k'} U(k,k') e^{i(k+k')x}.
\end{equation}
The energy eigenvalue in eqs. (\ref{ev1}) and (\ref{ev2}) is given by
\begin{equation}
E^{(2)}(k,k')=\omega(k)+\omega(k') \label{1+1disp},
\end{equation}
which is determined from the free part of the two DW problem.

We define the $S$-matrix of the two DW as $S(k,k')\equiv B/A$. Solving the scattering problem (\ref{ev1}) and (\ref{ev2}), we obtain 
\begin{eqnarray}
S(k,k')=
-e^{i(k'-k)}
\frac{e^{-i3k}+e^{i3k'}-\left[2(\cos( 3k)+\cos(3k'))-\alpha ( e^{i3(k+k')}+e^{-i3(k+k')})\right] }{e^{i3k}+e^{-i3k'}-\left[2(\cos(3k)+\cos(3k'))-\alpha ( e^{i3(k+k')}+e^{-i3(k+k')})\right]}\;. \label{upsmatrix}
\end{eqnarray}
\end{widetext}
In this expression of the $S$-matrix, the extra phase factor $e^{i(k'-k)}$ means that  if a DW overtakes another one, the spin array around the DW is shifted by one site.
This is a peculiar factor for the DW problem, unlike to the usual spin wave one.

We now consider the formation of the bound state in the two DW problem; we assume 
\begin{equation}
k=u+iv \quad {\rm and} \quad k'=u-iv \label{bkpair}
\end{equation}
which corresponds to the 2-string solution in the Bethe ansatz terminology.
Substituting eq.(\ref{bkpair}) into eq.(\ref{1+1disp}), we obtain the dispersion curve of the bound state particle
\begin{equation}
E^{\rm bound}(u)=\frac{2}{3}(J_1+J_2)+ \varepsilon J_1 (e^{-3v}+ e^{3v})\cos 3u. \label{bsdisp}
\end{equation}

From the normalizability condition of the wave function, $u$ and $v$ must satisfy a nontrivial condition:
\begin{equation}
-e^{3v} \cos(3u)+ 2 \alpha \cos^2 (3u) -\alpha =0 ,\label{bseq}
\end{equation}
which is equivalent to the pole of the $S$-matrix (\ref{upsmatrix}).
The formal solution of eq.(\ref{bseq}) is given by
\begin{equation}
e^{3v}=\alpha \frac{\cos 6u}{\cos 3u}. \label{bsfs}
\end{equation}

\begin{figure}[bt]
\epsfig{file=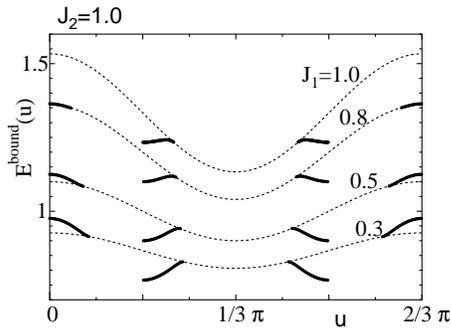, width=6cm}
\caption{The DW bound state dispersion curve for $\varepsilon=0.1$ and $J_2=1$.
The curves in the low-energy region are shown for $0<u<2\pi/3$.
The solid lines indicate the curves for $J_1=$1.0, 0.8, 0.5($= 1/\alpha^*_+$) and 0.4.
The dashed lines are the corresponding lower bounds of the free two DWs.
} \label{figupdisp}
\end{figure}

Since the physical solution must satisfy  $e^{3v}\ge 1$ and $-1 \le \cos 3u\le 1$,  the bound state dispersion curve emerges in the restricted range of $u$.
We  numerically evaluate eq.(\ref{bsfs}) in the permitted range of $u$ with $\varepsilon=0.1$, and illustrate the obtained dispersion curve (\ref{bsdisp}) in FIG.\ref{figupdisp}.
In the figure, we can see that the bound state band appears below the dispersion curve of two free DWs that is obtained as eq. (\ref{1+1disp}) with $k=k'$.

On the bases of figure \ref{figupdisp}, we discuss features of $E^{\rm bound}(u)$ in detail.
The low-energy solution of eq.(\ref{bsfs}) exists in the range 
\begin{equation}
\frac{\pi}{6} < u < u^*\quad {\rm or } \quad \frac{2\pi}{3}-u^* < u <\frac{\pi}{2},
\end{equation}
where $u^* = \frac{1}{3}\arccos [ \frac{1}{4}\left(\alpha^{-1}-\sqrt{\alpha^{-2}+8}\right)]$.
In addition to this, we note that, when $\alpha > 1$, eq.(\ref{bseq}) has a solution in the high energy region around $u\sim 0$ and $2\pi/3$.

In FIG.\ref{figupdisp}  we can see that, as $\alpha$ increases, the bound state dispersion curve comes down to the lower-energy region. 
Since the minimum energy of the free two-particle dispersion (\ref{1+1disp}) is given by  $E_{\rm min}^{(2)}\equiv 2(J_1+J_2)/3-2\varepsilon J_1$  at $k=k'=\pi/3$,  the bound state excitation becomes relevant, if the minimum of the dispersion (\ref{bsdisp}) is lower than $E_{\rm min}^{(2)}$.
Substituting eq.(\ref{bsfs}) into eq.(\ref{bsdisp}), we obtain the range of $\alpha$ where the bound state is relevant in the low-energy region:
\begin{equation}
 \alpha \ge 2 \;(\equiv \alpha^*_+). 
\end{equation}
This value of $\alpha^*_+$ is particularly important for the even-odd behavior of the magnetization curve, which is discussed in Sec. V. 

In addition, we note that the minimum energy of the DW bound state is achieved at $u \simeq \pi/6$ or $\pi/2$, where $\cos 3u \simeq 0$.
Thus we can easily see $E^{\rm bound}(u) \simeq 2(J_1+J_2)/3+ J_2 \cos 6u$ near the bottom of the bound-state dispersion curve.
This implies that the low-energy excitation for $J_2\gg J_1$ is described by the hopping of the DW bound state originating from the NNN term, which is consistent with the basic view of the DW bound state discussed at eq.(\ref{updwhop}).

\subsection{two body problem for the $S^z=-1/3$ DW}

We next consider the two-body problem for the $S^z=-1/3$ DWs.
Within the first order perturbation, we can solve the two DW problem with the same manner as the previous $S^z=1/3$ case. 

The single DW dispersion curve is obtained as
\begin{equation}
\omega(k)= \frac{-J_1+2J_2}{3} + \varepsilon J_1 \cos 3k
\end{equation}
The $S$-matrix of the two-DW problem is also calculated as
\begin{widetext}
\begin{eqnarray}
S(k,k')=
-e^{i(k-k')}
\frac{e^{-i3k}+e^{i3k'}-\left[2\bar{\Delta}+2(\cos( 3k)+\cos(3k'))-\alpha ( e^{i3(k+k')}+e^{-i3(k+k')})\right] }
{e^{i3k}+e^{-i3k'}-\left[2\bar{\Delta}+2(\cos(3k)+\cos(3k'))-\alpha ( e^{i3(k+k')}+e^{-i3(k+k')})\right]} \label{downsmatrix}
\end{eqnarray}
\end{widetext}
where 
\begin{equation}
\bar{\Delta}\equiv \frac{-1+2\alpha}{\varepsilon}.
\end{equation}
This $\bar{\Delta}$ terms originates form the ``binding energy'' in the zeroth order(Ising limit). 

\begin{figure}
\epsfig{file=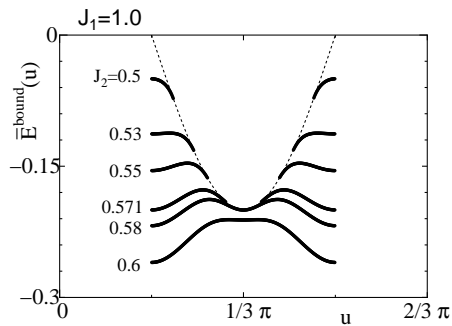, width=6cm}
\caption{The DW bound state dispersion curve of $\varepsilon=0.1$ around $\alpha=1/2$. 
The curves in the low-energy region are shown for $0<u<2\pi/3$.
The solid lines indicate the curves for $J_2=$0.5, 0.53, 0.55, 0.571($\simeq \alpha^*_-$), 0.58 and 0.6.
The dashed line is the lower bound of the two free DW dispersion.
In this figure, the origin of the energy axis is shifted with $\bar{E}^{\rm bound}= E^{\rm bound}+2(J_1-2J_2)/3$.}
\label{figdowndisp}
\end{figure}

Assuming the 2-string type solution (\ref{bkpair}), we obtain the equation for  the DW bound state:
\begin{equation}
-e^{3v}\cos 3u +2\alpha \cos^2 3u -\alpha-\bar{\Delta}=0,\label{dwnbseq}
\end{equation}
and its formal solution:
\begin{equation}
e^{3v}= \frac{ \alpha  \cos 6u -\bar{\Delta}}{\cos 3u}.\label{dwnbsev}
\end{equation}
The dispersion curve of the DW bound state is
\begin{equation}
E^{\rm bound}(u)=\frac{2}{3}(-J_1+2J_2)+ \varepsilon J_1(e^{-3v}+ e^{3v})\cos 3u. \label{dwnbsdisp}
\end{equation}
Although this expression of the dispersion looks very similar to the $S^z=2/3$ case, the resulting behavior of the curve is quite different;
Since the $\bar{\Delta}$ contains the factor $1/\varepsilon$, the dominant properties of (\ref{dwnbsev}) is well approximated to be $e^{3v}\simeq -\bar{\Delta}/\cos 3u$ as far as $|\bar{\Delta}|>\varepsilon $.
However, for $\bar{\Delta}<\varepsilon$, namely, in the vicinity of $\alpha=1/2$,  the curve depends on $\alpha$ sensitively.
We thus show the bound state dispersion (\ref{dwnbsdisp}) of $\varepsilon=0.1$ around $\alpha =1/2$ in FIG.\ref{figdowndisp}, where we can actually see that the behavior of the curve changes drastically.

When $\alpha <\frac{1-\varepsilon}{2-\varepsilon}$, the solution of eq.(\ref{dwnbseq}) appears only in the high energy region($u\simeq 0$).
As $\alpha$ increases, the bound state excitation is able to appear in the low-energy region; For  
\begin{equation}
\alpha > \frac{1}{2+\varepsilon},
\end{equation}
the bound state dispersion curve appears at $u = \pi/6$ and $\pi/2$.
As $\alpha$ is increased beyond this value, the two branches of the curve are extended from the band edges toward $u=\pi/3$.
At the same time, the energy at  $u = \pi/6$ and $\pi/2$ decreases rapidly.
We can see that $E^{\rm bound}(u)$ has the local minimum at the band edges for 
\begin{equation}
\alpha > \frac{1+\sqrt{1+2\varepsilon(2+\varepsilon)}}{2(2+\varepsilon)}(\equiv \tilde{\alpha}_-),
\end{equation}
where $\tilde{\alpha}_-$ is determined by $\frac{d^2 E^{\rm bound}(u)}{d u^2}|_{u=\pi/6 \;{\rm or}\;\pi/2} =0$. 

Since the minimum value of the two free DW dispersion is given by $E_{\rm min}^{(2)}\equiv 2(-J_1+2J_2)/3-2\varepsilon J_1$,
we can further see that the bottom of the bound-state dispersion curve becomes lower than $E_{\rm min}^{(2)}$, when 
\begin{equation}
\alpha >  \frac{1+2\varepsilon}{2+\varepsilon} \; (\equiv \alpha^*_-).
\end{equation}
This value of $\alpha^*_-$ is also important for the  analysis of the even-odd behavior of the magnetization curve.

As $\alpha$ is increased further,  after the branches extended from $u=\pi/6$ and $\pi/2$ connect with each other at $\alpha=\frac{1+\varepsilon}{2-\varepsilon}$,  $E^{\rm bound}(u)$ approaches $\frac{1}{3}(J_1-2J_2)+\varepsilon J_2 \cos 6u$ rapidly.
This behavior of $E^{\rm bound}(u)$ is also consistent with the DW bound state  moving with use of the NNN term when $J_2\gg J_1$.

\section{magnetization curve}

In order to verify the analytical results for the DW excitations around the 1/3 plateau, we calculate the magnetization curve for the Ising-like zigzag XXZ chain with the DMRG method.
We discuss the relevance of the fractional $S^z$ excitations to the obtained magnetization curves.

\subsection{numerical results}

We calculate the magnetization curve for the zigzag XXZ chain of 192 sites($N=64$) for $\varepsilon=0.1$, using the DMRG method.
The number of the retained bases used in the DMRG computations is typically 64, with which the calculated magnetization curves converges sufficiently.

\begin{figure}[ht]
\epsfig{file=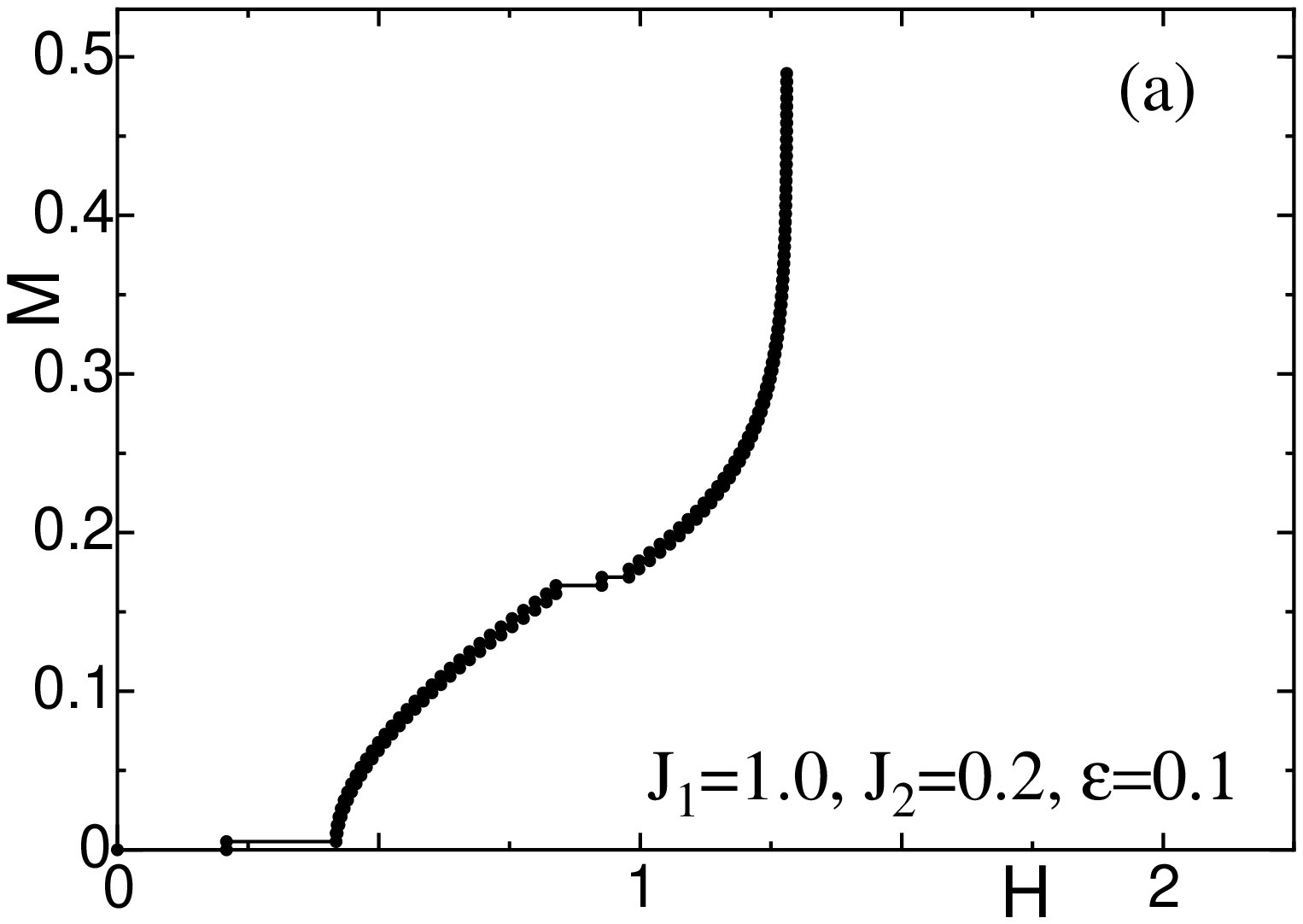, width=6cm}
\epsfig{file=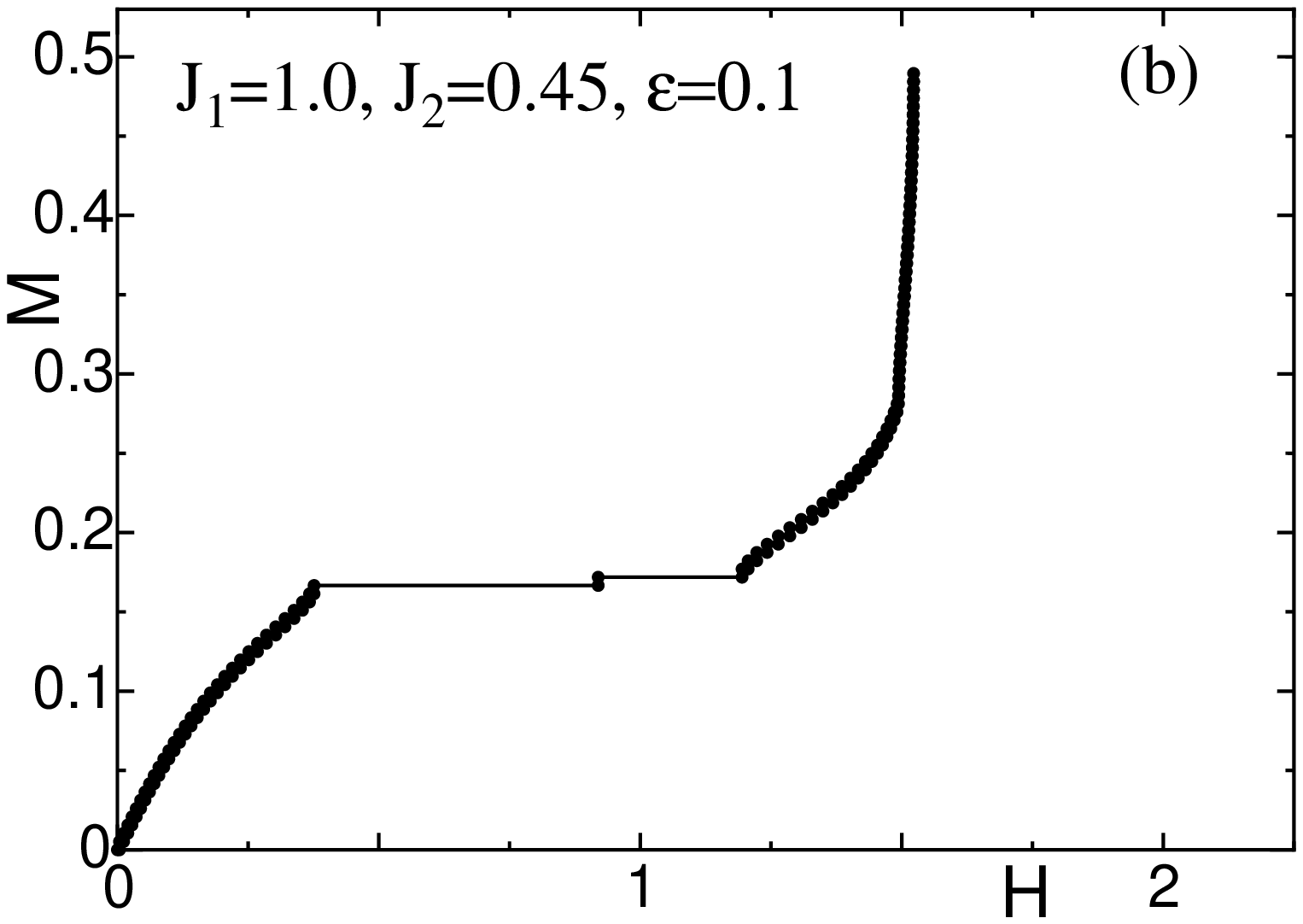, width=6cm}
\epsfig{file=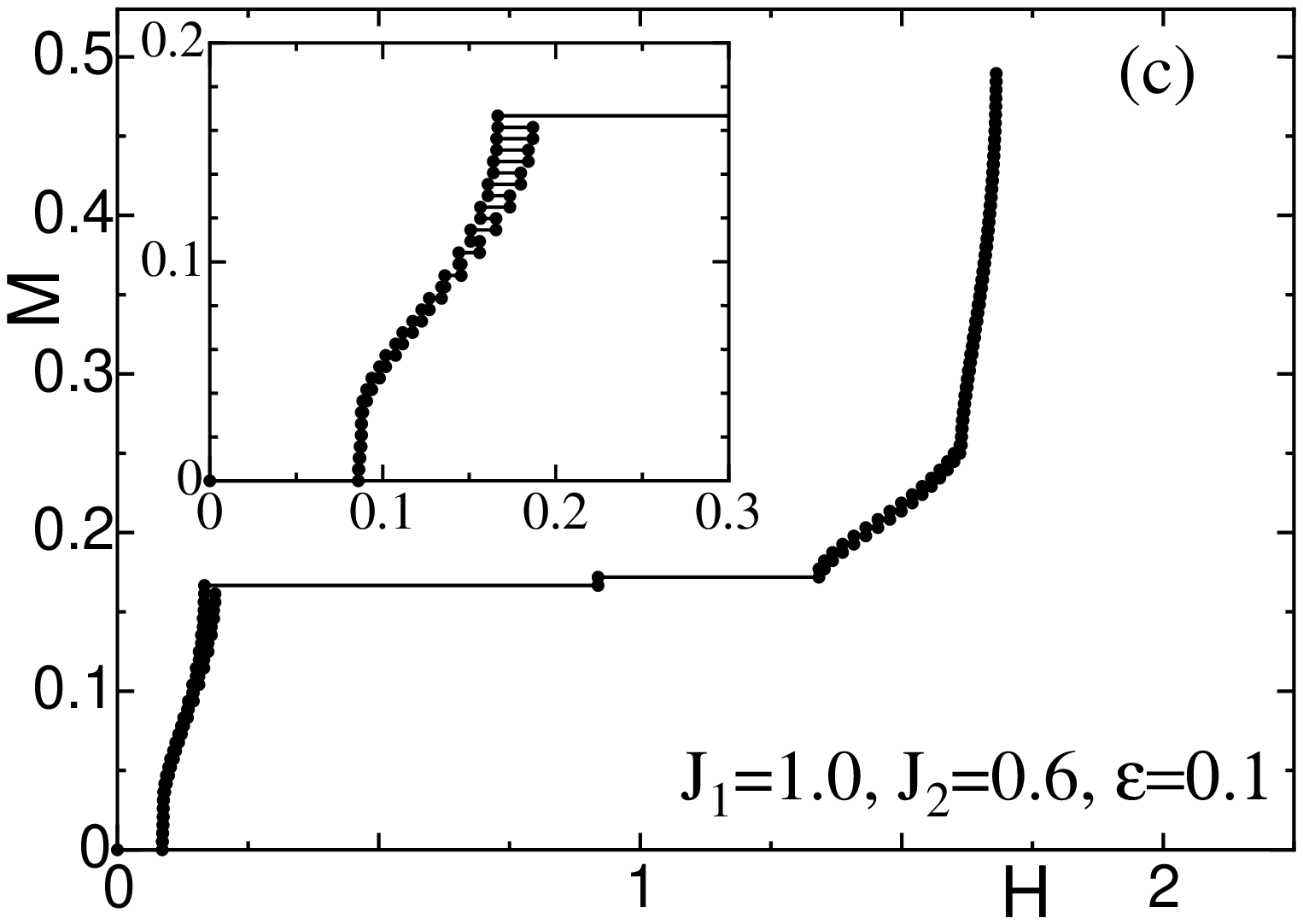, width=6cm}
\epsfig{file=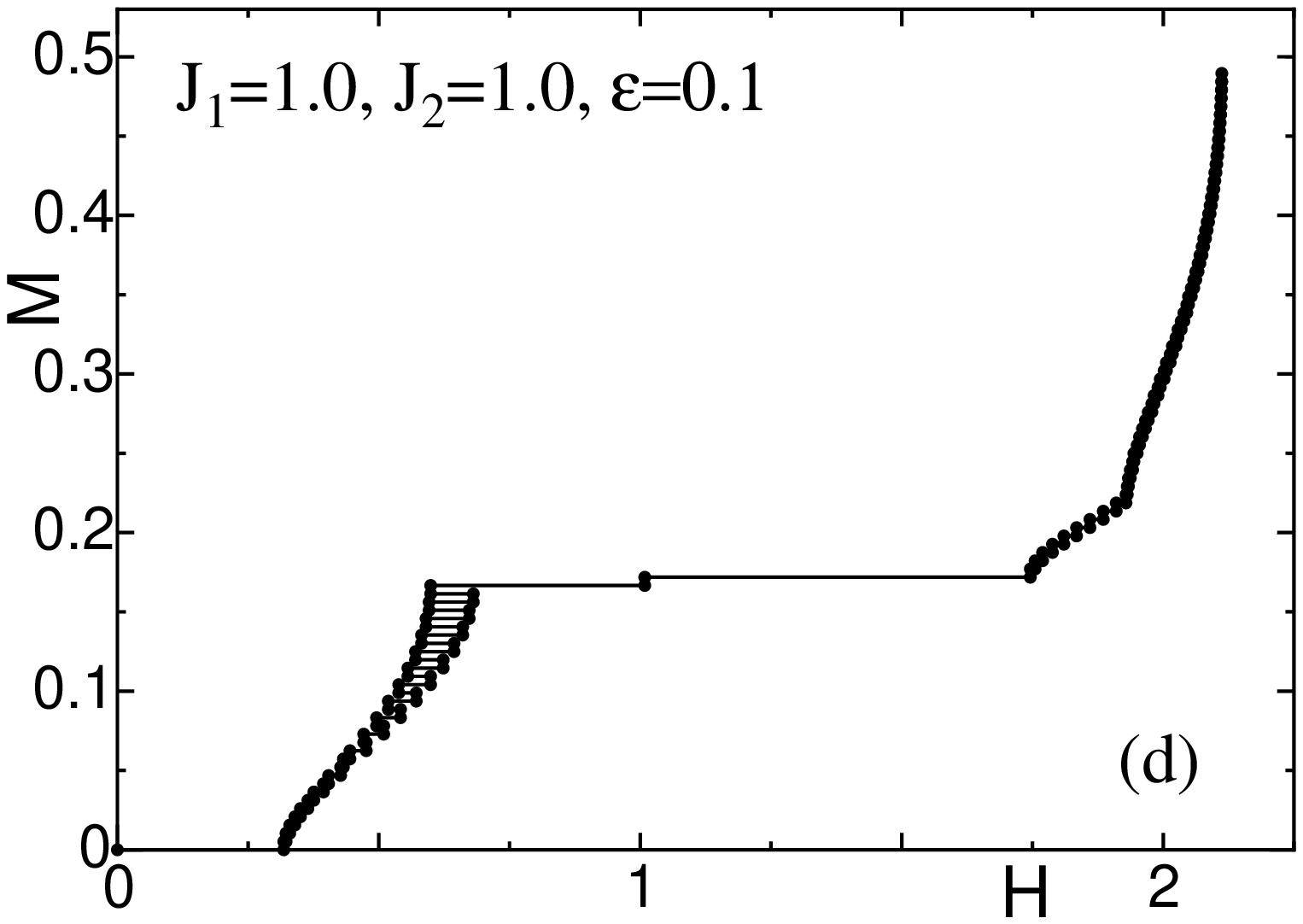, width=6cm}
\caption{The magnetization curves of the Ising-like zigzag XXZ chain of the 192 sites for $\varepsilon=0.1$. (a)$J_2=0.2$, (b)0.5, (c)0.6 and (d)1.0.
The inset in figure (c) is the magnification around $H\simeq 0.2$, where the even-odd oscillation appears in the low-field branch.}
\label{mh1}
\end{figure}

Let us first survey the features of the curves.
In FIG.\ref{mh1}, we show the magnetization curves for $J_1\ge J_2$, where we fix $J_1=1$ and vary $J_2$.
As $J_2$ is increased from $J_2=0$, the magnetization plateau emerges at 1/3 of the full moment(FIG.\ref{mh1}-(a)), and the width of the 1/3 plateau extends rapidly.
At the same time, the gap at zero magnetic field becomes small.
This is consistent with the transition form the N\'eel($\uparrow\downarrow$) phase to the double-N\'eel($\uparrow\uparrow\downarrow\downarrow$) one in the Ising limit, which is  located at $\alpha =1/2$ and $H=0$. 
A precise study of the critical point of the N\'eel and double-N\'eel transition for the zigzag XXZ model can be seen in Ref.\cite{okamoto}.
The estimated value of the critical point for $\varepsilon=0.1$ is  $J_2\simeq 0.45$, which is in good agreement with the magnetization curve in FIG.\ref{mh1}-(b).
In addition to this, we can see that the low-field branch(= the magnetization curve below the 1/3 plateau) becomes very narrow.
As was mentioned in Sec. II, the zigzag Ising chain is highly degenerating at $\alpha=1/2$ below the 1/3 plateau, where the phase boundary of the 1/3 plateau also merges with the N\'eel and double-N\'eel transition point(see FIG.\ref{ilimit}).
For the XXZ chain of $\varepsilon\ne 0$, the quantum fluctuation coming from the XY term lifts this degeneracy, and thus the magnetization curve appears in the narrow region between the zero magnetization and the 1/3 plateau.
The detailed relation of the DW excitation and the magnetization curve is discussed in the next subsection.

Turning to the high-field branch of the magnetization curve, we can see that the cusp singularity  appears in the high-field branch for $J_2>1/4$, which can be explained well by the shape change of the spin-wave dispersion curve from the saturated state.\cite{cusp}

In FIG.\ref{mh1}-(c), we can find that the even-odd oscillation appears in the low-field branch.
Although the true magnetization curve of the finite size system should be determined from the stable $M=$even states,  we here show the anomalous $M=$odd steps as well, because it is the evidence of the DW bound state having $S^z=\pm 2/3$.
We discuss the detailed mechanism of the even-odd oscillation based on the DW picture in the next subsection. Here we note that the even-odd behavior emerges for $J_2 \gsim 0.58$.
As $J_2$ is increased further, the amplitude of the even-odd oscillation in the low-field branch is extended gradually, and at the same time the cusp singularity in the high-field branch shifts to the low magnetization side(see FIG.\ref{mh1}-(d)).

\begin{figure}[ht]
\epsfig{file=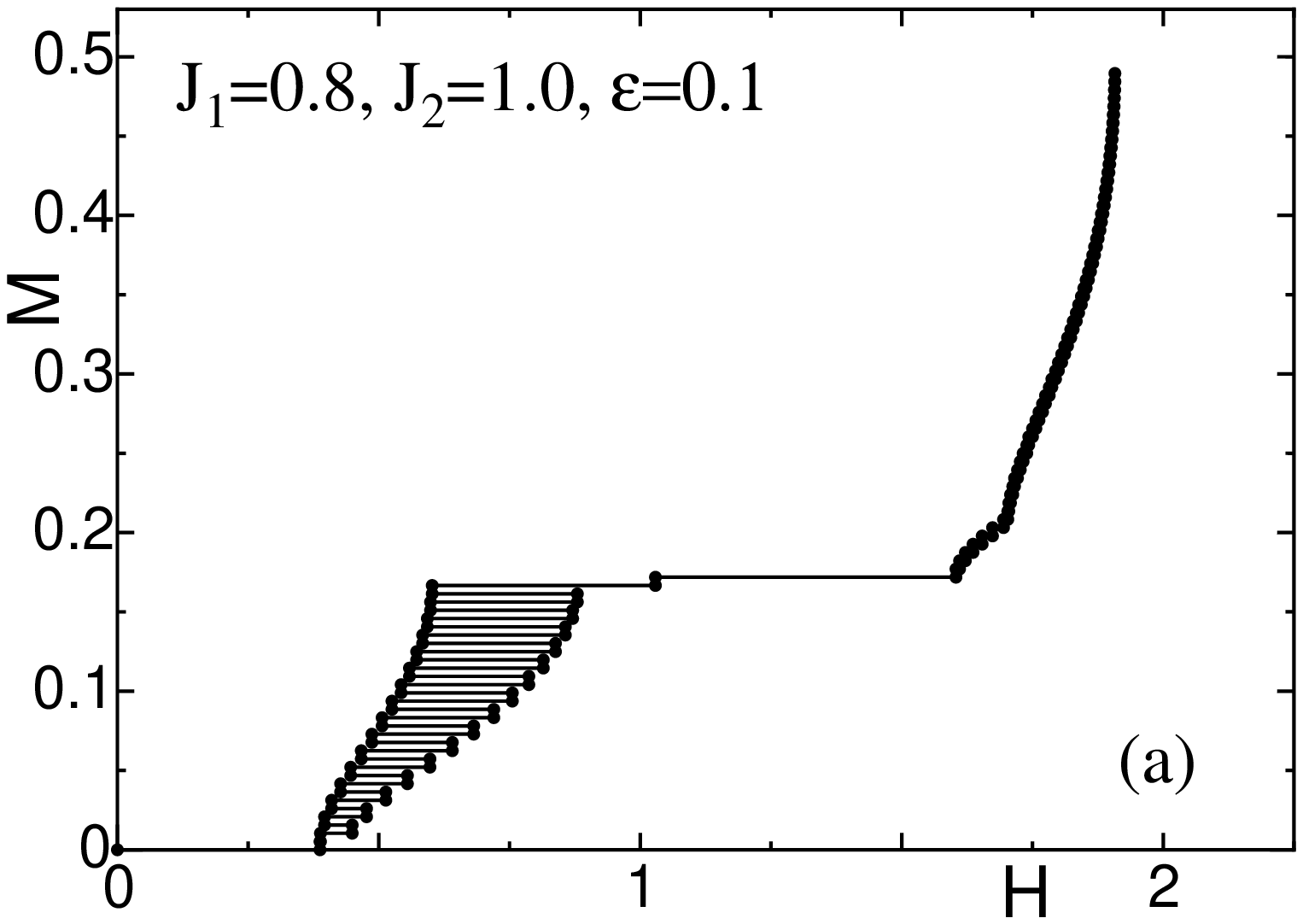, width=6cm}
\epsfig{file=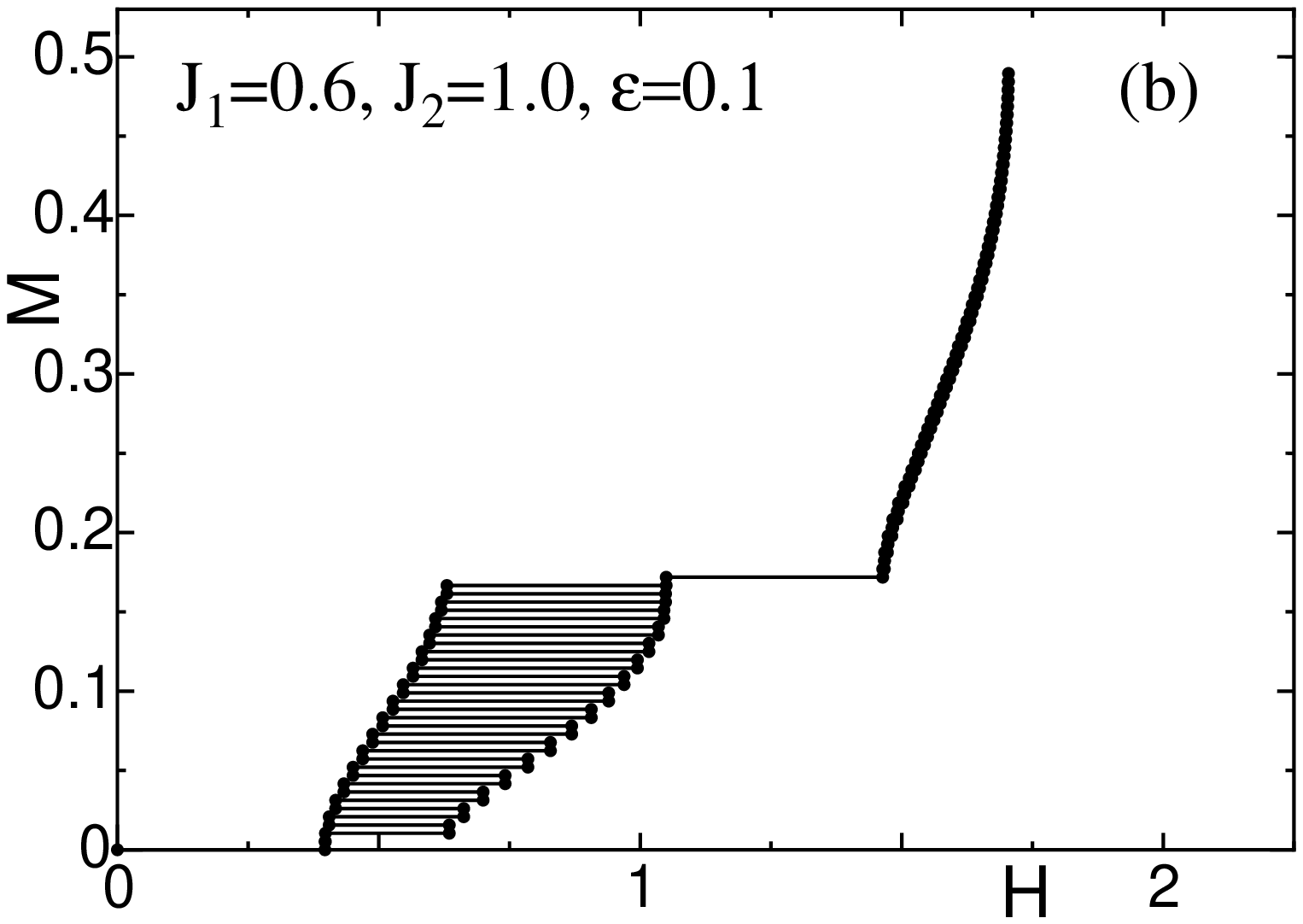, width=6cm}
\epsfig{file=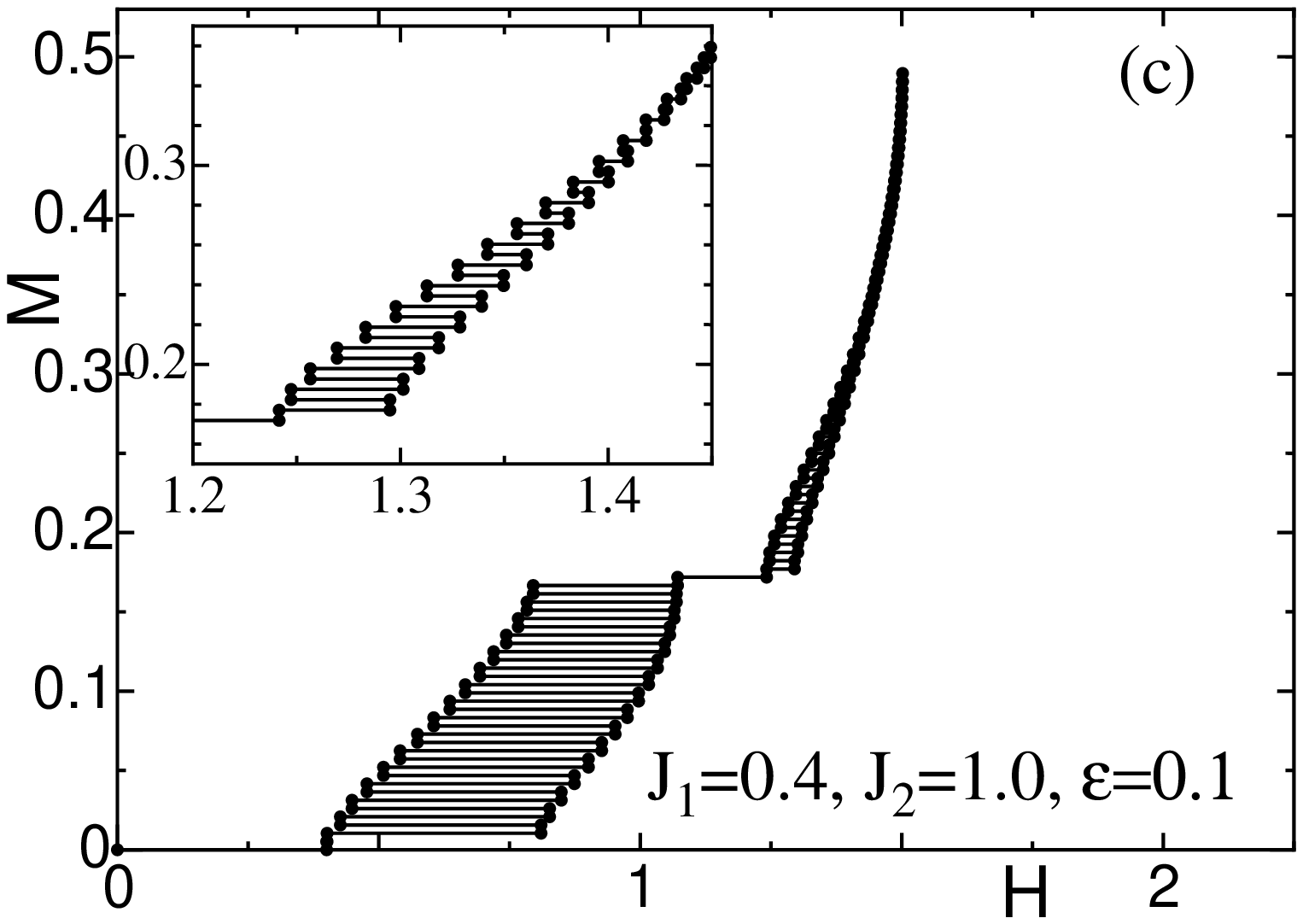, width=6cm}
\caption{The magnetization curves of the Ising-like zigzag XXZ chain of the 192 sites for $\varepsilon=0.1$. (a)$J_1=0.8$, (b)0.6, and (c)0.4.
The inset in figure (c) is the magnification around $H\simeq 1.3$, where the even-odd oscillation appears in the high-field branch.}
\label{mh2}
\end{figure}

We further show the magnetization curves for $J_1 <J_2$ in FIG.\ref{mh2}.
In the figures we fix $J_2=1$ and vary $J_1$.
As $J_1$ is decreased from unity, the width of the 1/3 plateau shrinks. 
Simultaneously  the region of the even-odd oscillation in the low-field branch extends down to the zero magnetization, and the amplitude of the oscillation becomes significant(FIG.\ref{mh2}-(a)). 
On the other hand,  for the high field branch, the position of the high-field cusp approaches the 1/3 plateau;
Indeed for the magnetization curve of $J_1\simeq 0.6$ in FIG.\ref{mh2}-(b) we can see that the high-field cusp merges into the 1/3 plateau.
For $J_1 < 0.6$, the cusp singularity does not appear, but, instead, the even-odd oscillation emerges in the high-field branch.
In FIG.\ref{mh2}-(c), we can confirm the oscillating behavior of the magnetization curve above the 1/3 plateau clearly.
In $J_1\to 0$ limit, the 1/3 plateau disappears and the magnetization curve finally becomes that of the two decoupled  XXZ spin chain.
Here it should be remarked that, in the high-field region, the phase diagram of  the Ising limit contains no special point having a peculiar degeneracy, unlike to the previous N\'eel and double-N\'eel transition point in the low-field case.
This suggests that the even-odd oscillation in the high-field branch is induced by the purely quantum effect.

\subsection{DW excitation and the magnetization curve}

We now discuss the relevance of the DW excitations to the magnetization curves.
The magnetization curve of the 1D quantum spin systems is generally described  by the hard-core bosonic particle picture;\cite{sorensen,oha}
a particle having magnetization fills its dispersion curve up to the ``chemical potential'' corresponding to the external magnetic field.
Then the magnetization curve is interpreted as the chemical potential v.s. the particle-number curve, where the shape of the dispersion curve are  essentially important to figure out  the feature of the magnetization curve.
As was seen in Sec. III, the elementally excitations around the 1/3 plateau are described by the DW particles carrying $S^z= \pm 1/3$ and their bound states having $S^z=\pm 2/3$, although the total magnetization of the system  always takes  an integer value particularly for a finite size system.
This implies that the magnetic excitation of $S^z=\pm 1$ around the plateau state is described by the combinations of the fractional $S^z$ DW excitations.

Let us first discuss the relation between the $S^z=+1/3$ excitation and the characteristic properties of the high-field branch.
If $J_1\gg J_2$, the effect of NNN term is not so big.
Thus the low-energy excitation around the plateau state is basically described by the free $S^z=1/3$ DW excitation.
The magnetization curve near the 1/3 plateau is reflecting the shape of the $\omega(k)$ for the  $S^z=1/3$ DW;
the magnetization curve raises from the 1/3 plateau with the square-root behavior associated with the curvature around the bottom of $\omega(k)$.\cite{second}
This is the case for $J_2< 1/4$ in the high field branch(FIG.\ref{mh1}-(a))

\begin{figure}
\epsfig{file=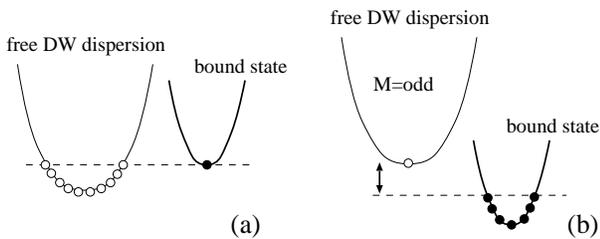, width=8cm}
\caption{Schematic diagram (a) for the cusp singularity and (b) for the even-odd oscillation of the magnetization curve. The open circle denotes the DW excitation and the solid circle denotes the DW bound state. The broken lines indicate the ``chemical potential'' corresponding to the magnetic field}
\label{schem1}
\end{figure}

As $J_2$ is increased and the NNN effect becomes more significant,  the dispersion curve of the DW bound state comes down to the low-energy region.
Then there are two possible situations: the bottom of the DW-bound-state dispersion curve is slightly higher than that of the single DW dispersion curve(FIG.\ref{schem1}-(a)), and the opposite (FIG.\ref{schem1}-(b))

For the former case, the magnetization increases along the single DW dispersion curve, as long as  the chemical potential is below the bottom of the bound state dispersion curve. 
However, when the chemical potential  touches the bottom of the bound-state dispersion curve,  the magnetization curve captures the band edge singularity to have the cusp(FIG.\ref{schem1}-(a)).
After the chemical potential exceeds the bottom of the bound-state dispersion curve, the magnetization can increases rapidly by using the bound-state dispersion curve.
In the context of the crossing points of the chemical potential and the dispersion curves, the above obserbation implies that the two component TL liquid is realized above the cusp singularity, while the one component TL liquid is realized between the high-field cusp and the 1/3 plateau.
These behaviors of the each branch of the magnetization curve is consistent with the known results based on the shape change of the spin-wave dispersion curve from the saturation limit\cite{cusp}.
The situation illustrated in FIG.\ref{schem1}-(a) corresponds to  the case  for $\alpha <\alpha^*_+$.

For the latter case, the magnetization curve increases by using the bound state dispersion curve.
We here recall that the excitation of $S^z=1$ is always represented as a combination of three DW particles.
If $M=$odd, there are $3(M-N)$ number of $S^z =1/3$ DW particles in the system.\cite{evensite}
Then  $3(M-N)-1$ numbers of the DWs can conform the bound state of $S^z=2/3$, but the remaining one DW can not find its partner.
Thus the remaining one DW have to sit on the free DW dispersion curve, as in the diagram of FIG.\ref{schem1}-(b).
Therefore the $M=$odd state has a slightly higher energy due to the gap between the bottom of the single DW band and the chemical potential lying in the bound state band.
On the other hand, for the case of $M=$even, all of the DWs can find their partners and conform the bound states successfully.
Clearly this is the origin of the even-odd behavior of the magnetization curve.

According to the analysis in Sec. IV,  we can see that the switching of the above two cases occurs at $\alpha=\alpha_+^*=2$.
Although the theoretical prediction $\alpha^{-1}=1/2$ is consistent with the DMRG result $\alpha^{-1}\simeq 0.6$, they shows a slight deviation.
The reason for this is that the $S^z=1/3$ DWs have  no ``binding energy''  in the zeroth order(Ising limit), corresponding to $\bar{\Delta}=0$ in eq. (\ref{upsmatrix}).
Thus the leading term for the binding energy is of first order of $\varepsilon$, implying that the formation of the DW bound state is responsible for the purely quantum effect.
For the precise determination of $\alpha^*_+$, the second order calculation with respect to $\varepsilon$ is required, where the application of the Bethe form wave function is rather difficult.

We turn to the analysis of the low-field branch, which can be explained by almost the same line of the argument.
However, in contrast to the high-field branch, the $S^z=-1/3$ DW excitation has the zeroth order binding energy, namely $\bar{\Delta}$ term,  which yields the sensitive behaviors of the dispersion curve of the DW bound state near $\alpha=1/2$.
Thus we first consider the case where $\alpha$ is sufficiently away from $1/2$, where the shape of the bound state dispersion is rather simple.
When  $\alpha <\frac{1}{2+\varepsilon}$, the bound state DW does not appear in the low energy region. 
Thus the magnetization curve can be described by the free $S^z=-1/3$ dispersion curve(FIG.\ref{mh1}-(a) and (b)).
On the other hand, if $J_2 > 0.62$, the shape of the bound state dispersion curve is approximated well by $E^{(2)}\simeq \frac{1}{3}(-J_1+2J_2)+ \varepsilon J_2 \cos 6k$, whose bottom is lower than that of the single DW dispersion.
Thus the low-field branch of the magnetization curve exhibits the even-odd oscillation for $J_2 > 0.62$(FIG.\ref{mh1}(d) and FIG.\ref{mh2}), due to the same mechanism in  FIG.\ref{schem1}-(b).

\begin{figure}
\epsfig{file=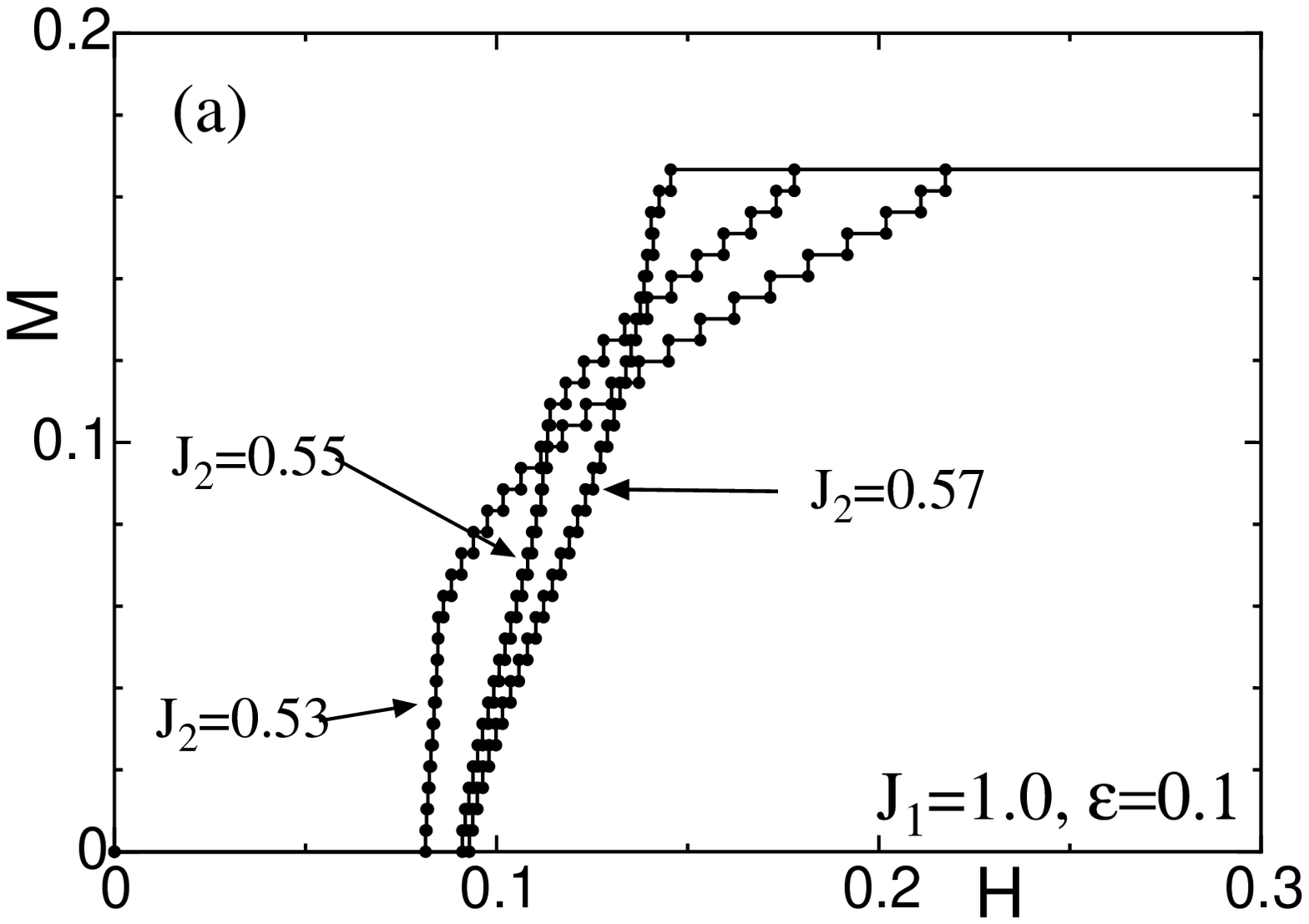, width=5.5cm}
\epsfig{file=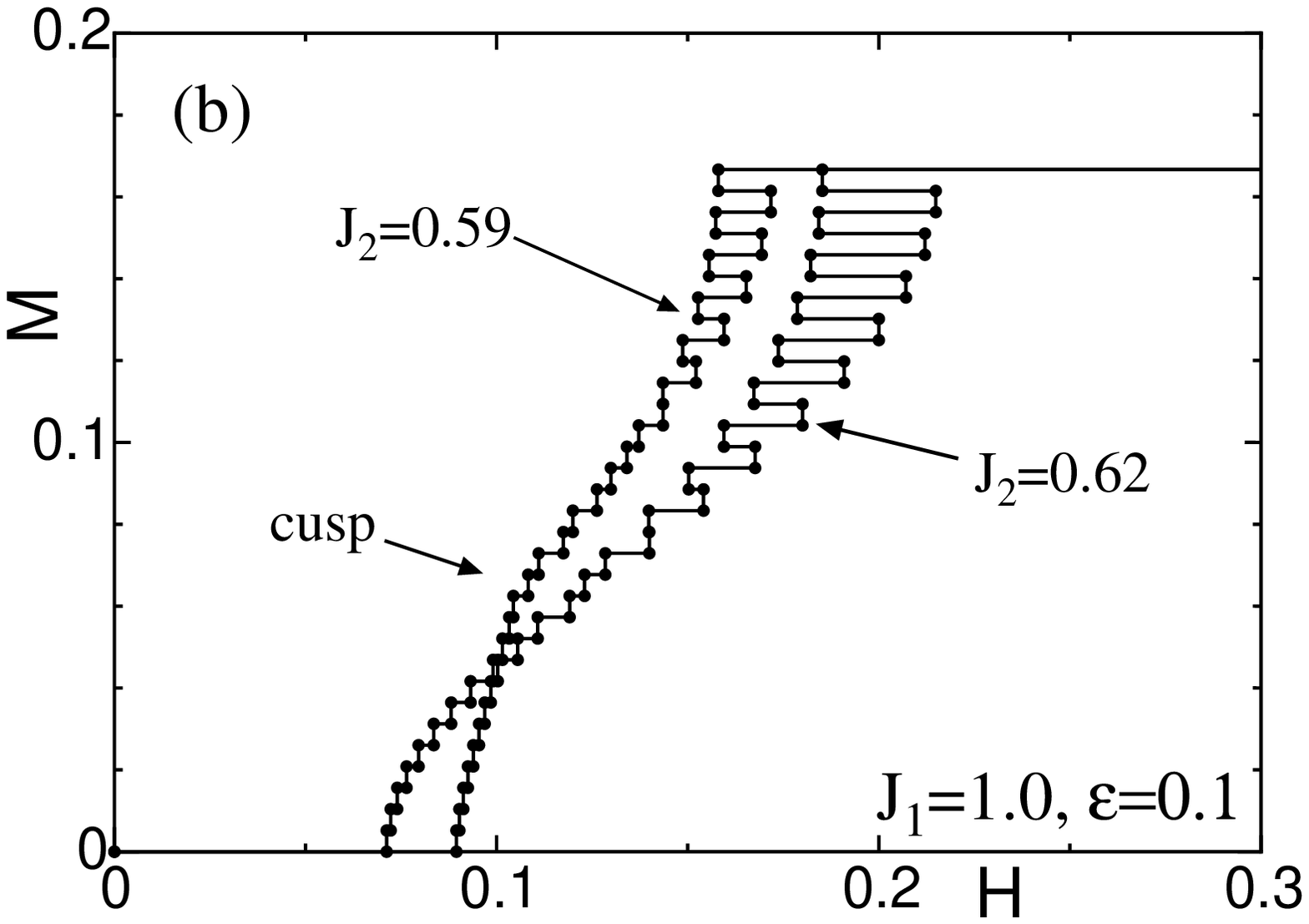, width=5.5cm}
\caption{The low-field branch of the zigzag XXZ chain near $\alpha=1/2$:
(a) $\alpha=0.53$, 0.55 and 0.57, and (b) $\alpha=0.59$ and 0.62.}
\label{near05}
\end{figure}

We next proceed to the analysis for $\alpha\simeq 1/2$.
As is seen in FIG.\ref{figdowndisp}, a characteristic feature of the $S^z=-2/3$ DW bound state is that the shape of the dispersion curve  changes  drastically in narrow range of $\alpha$.
Accordingly, we can expect that the shape of the low-field branch of the magnetization curve also changes in the corresponding region of $\alpha$.
In order to verify this expectation, we calculate the the low-field branch of the magnetization curves for $0.5<\alpha \le 0.62$ intensively, which are shown in FIG.\ref{near05}.
When $\alpha <\tilde{\alpha}_-=0.521$, the small dispersion curve of the bound state appears accompanying a local maximum at the band edges($u=\pi/6$ or $u=\pi/2$), which is located in a sufficiently higher energy region than the bottom of the free DW dispersion curve.
Thus the magnetization below the 1/3 plateau decreases using the free DW band.
For $\tilde{\alpha}_-<\alpha<\alpha^*_-=0.571$, the bound state dispersion comes down to the low energy region and has the local minimum at the band edges. 
Nevertheless the energy at this local minimum is still higher than the bottom of the free DW dispersion.
Thus, in this region of $\alpha$, the situation is the same as the case of FIG.\ref{schem1}-(a), so that the magnetization curve has the cusp singularity(FIG.\ref{near05}-(a)).

For $\alpha > \alpha^*_-$, the lowest energy of the bound state dispersion becomes lower than that of the single DW one, and thus the magnetization curve exhibits the even-odd oscillation, which is the case of FIG.\ref{schem1}-(b) and FIG.\ref{near05}-(b).
However, we should  notice that the bound state dispersion curves has another feature
 especially for  $0.579<\alpha<0.605;$\cite{fsolv}  the bound state dispersion curve has the {\it local  minimum at $u=\pi/3$},  implying that the curve shows the double-well shape.
Thus the magnetization curve may have the cusp singularity and even-odd behavior at the same time; the DMRG result for $\alpha = 0.59$ and 0.6 actually shows the both of the cusp and even-odd oscillation(FIG.\ref{near05}-(b), see also the inset of FIG.\ref{mh1}-(c)).
For $\alpha >0.605$, the bound state dispersion has the single minimum at $u=\pi/6$ or $\pi/2$, and then the magnetization curve exhibits the even-odd oscillation without the cusp.

As seen above, the DW excitation picture of fractional $S^z$ can explain well the magnetization curve calculated by the DMRG.
In the remaining part of this section,  we shall discuss to what extent  the present theory can retain its validity.
Since the dispersion curves in Sec.IV are obtained within the first order perturbation, the higher order contribution with respect to $\varepsilon$ may affect the shape of the dispersion curves.
However, what we want to emphasize here is  that the present DMRG results actually justifies the DW picture as far as $\varepsilon=0.1$.
Moreover, it should be noted that, even for the isotropic zigzag chain, the topology of the magnetic phase diagram is almost the same as the present Ising-like XXZ model, except for the dimer-gapped phase at the zero magnetic field of the isotropic case.
The fact of the 1/3 plateau existing for the isotropic zigzag chain suggests that the DW picture is basically maintained  against the quantum fluctuation due to the XY term.
Of course, a further precise analysis is required for the quantitative understanding of the problem especially around $\alpha \simeq 1/2$, which may be an interesting future subject.

Another important factor which should be added in the present theory is the interaction effect between the DWs.
Although the DMRG results also demonstrate that the DW picture basically holds up to the certain range of the magnetization curve, a careful consideration of the interaction between DWs is required for a more quantitative analysis for the magnetization curve away from the 1/3 plateau.
For example, the DWs and their bound state picture for the even-odd oscillation suggests that the magnetization curve may have some singularity at the upper edges of the even-odd oscillation branch, where the chemical potential touches the bottom of the free DW dispersion curve.
However we can not confirm such singularity within the DMRG result for the 192 sites chain.
A possible reason for this is the interaction effect, since the number of the $S^z=1/3 $ DWs increases as the magnetization is increased away from the 1/3 plateau.

\section{OBC and defect in the 1/3 plateau state}

\subsection{Ising limit}

In this section we give a remark about the plateau state for the OBC system.  
Since the DMRG gives the results for the OBC, the plateau state for the OBC has a special importance for the alanysis of the DMRG results.

For the case of the PBC, it is not permitted to insert single DW into the plateau state without changing the total magnetization.
In other words, the low-energy  excitation of the $M=N$ sector is always represented as a combination of two DWs such as the pair-creation of the $S^z=\pm1/3$ particles. 
Thus the excitation that does not change the total $S^z$ of the system always accompanies the energy rise of order $J_1$ or $J_2$.

For the case of the OBC, however, we can insert single DW into the plateau state with keeping the magnetization of the system.
Thus we should check the energy of not only the ``uniform N\'eel state'' but also the state inserted the single DW. 
The three uniform N\'eel states for the OBC are:
\begin{eqnarray*}
|\widetilde{\rm N\acute{e}el}_1\rangle&=& \uparrow\downarrow\uparrow\uparrow\downarrow\uparrow\cdots\uparrow\downarrow\uparrow\uparrow\downarrow\uparrow \\
|\widetilde{\rm N\acute{e}el}_2\rangle&=& \uparrow\uparrow\downarrow\uparrow\uparrow\downarrow\cdots\uparrow\uparrow\downarrow\uparrow\uparrow\downarrow \\
|\widetilde{\rm N\acute{e}el}_3\rangle&=& \downarrow\uparrow\uparrow\downarrow\uparrow\uparrow\cdots\downarrow\uparrow\uparrow\downarrow\uparrow\uparrow \, ,
\end{eqnarray*}
and the corresponding energies are easily obtained as
\begin{equation}
\tilde{E}_1= E_g - \frac{J_1}{4} +\frac{J_2}{2}, \qquad \tilde{E}_2=\tilde{E}_3=E_g + \frac{J_1}{4}. \label{uniopen}
\end{equation}
Since the DWs of $S^z=-1/3$ or $-2/3$ have ``negative energy'' depending on the coupling constants,  we should  evaluate the energy for the following three states:
\begin{eqnarray*}
|\phi_{13}\rangle&=& \uparrow\downarrow\uparrow\cdots\uparrow\downarrow\uparrow\downarrow\uparrow\uparrow\cdots \downarrow\uparrow\uparrow \\
|\phi_{21}\rangle&=& \downarrow\uparrow\uparrow\cdots\uparrow\uparrow\downarrow\uparrow\downarrow\uparrow\cdots \uparrow\downarrow\uparrow \\
|\phi_{23}\rangle&=& \uparrow\uparrow\downarrow\cdots\uparrow\uparrow\downarrow\downarrow\uparrow\uparrow\cdots \downarrow\uparrow\uparrow ,
\end{eqnarray*}
which may have lower energies than those of the uniform N\'eel states (\ref{uniopen}.
The energies for these states are easily calculated as
\begin{eqnarray}
& &\tilde{E}_{13}=\tilde{E}_{21}=E_g-\frac{J_1}{4}+J_2, \\
& &\tilde{E}_{23}=E_g+\frac{3J_1}{4}-\frac{J_2}{2}.
\end{eqnarray}
Thus we can conclude that the 1/3 plateau state for OBC is $|\widetilde{\rm N\acute{e}el}_1\rangle$ for $ J_2 < J_1$ and $ |\phi_{23}\rangle $ for $J_2 > J_1$.
Although this is the case of the Ising limit, we will see that the 1/3 plateau state that is inserted the DW survives against the quantum fluctuation.

\subsection{DMRG result}

\begin{figure}[ht]
\epsfig{file=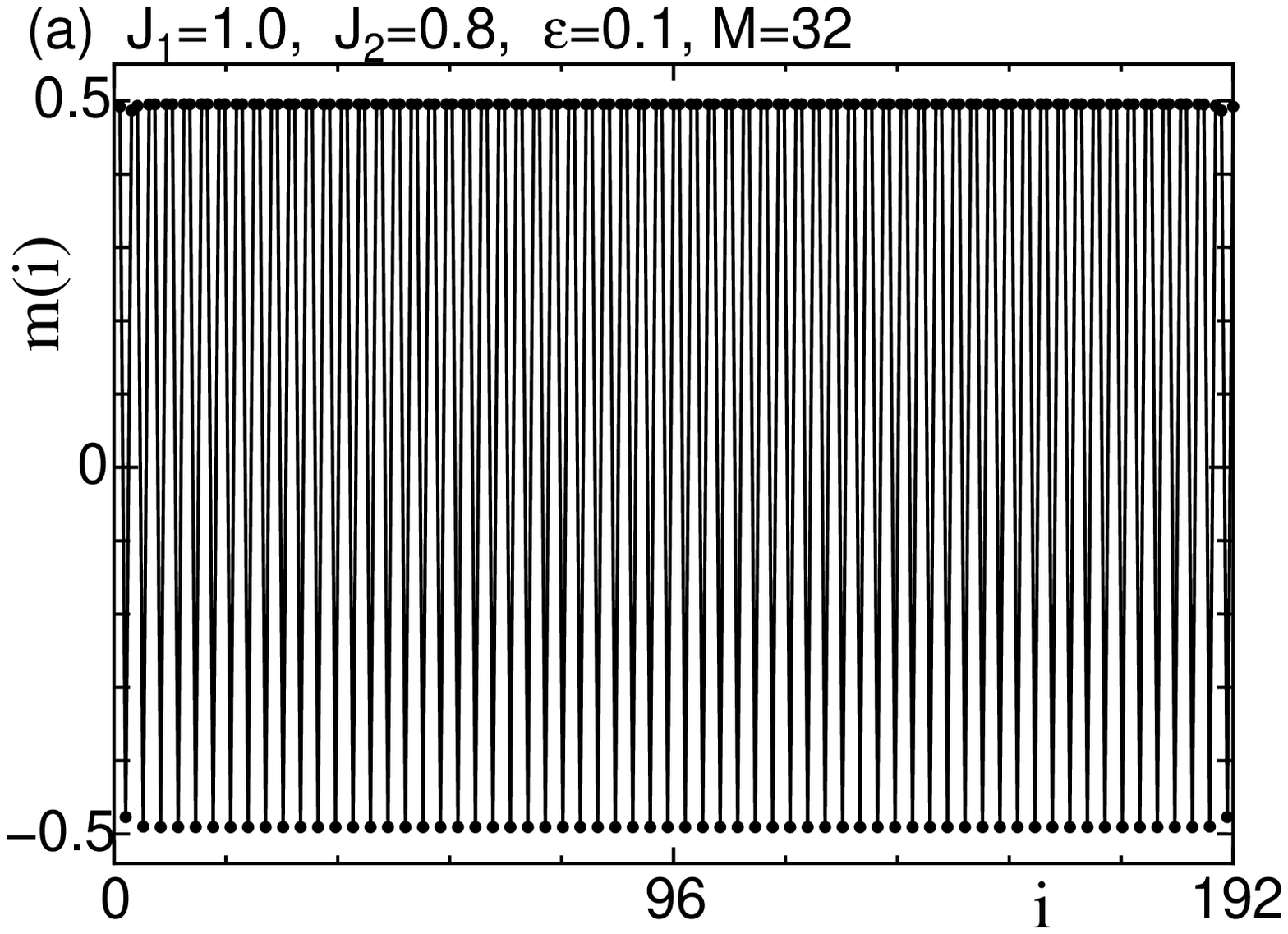,width=5.5cm}
\epsfig{file=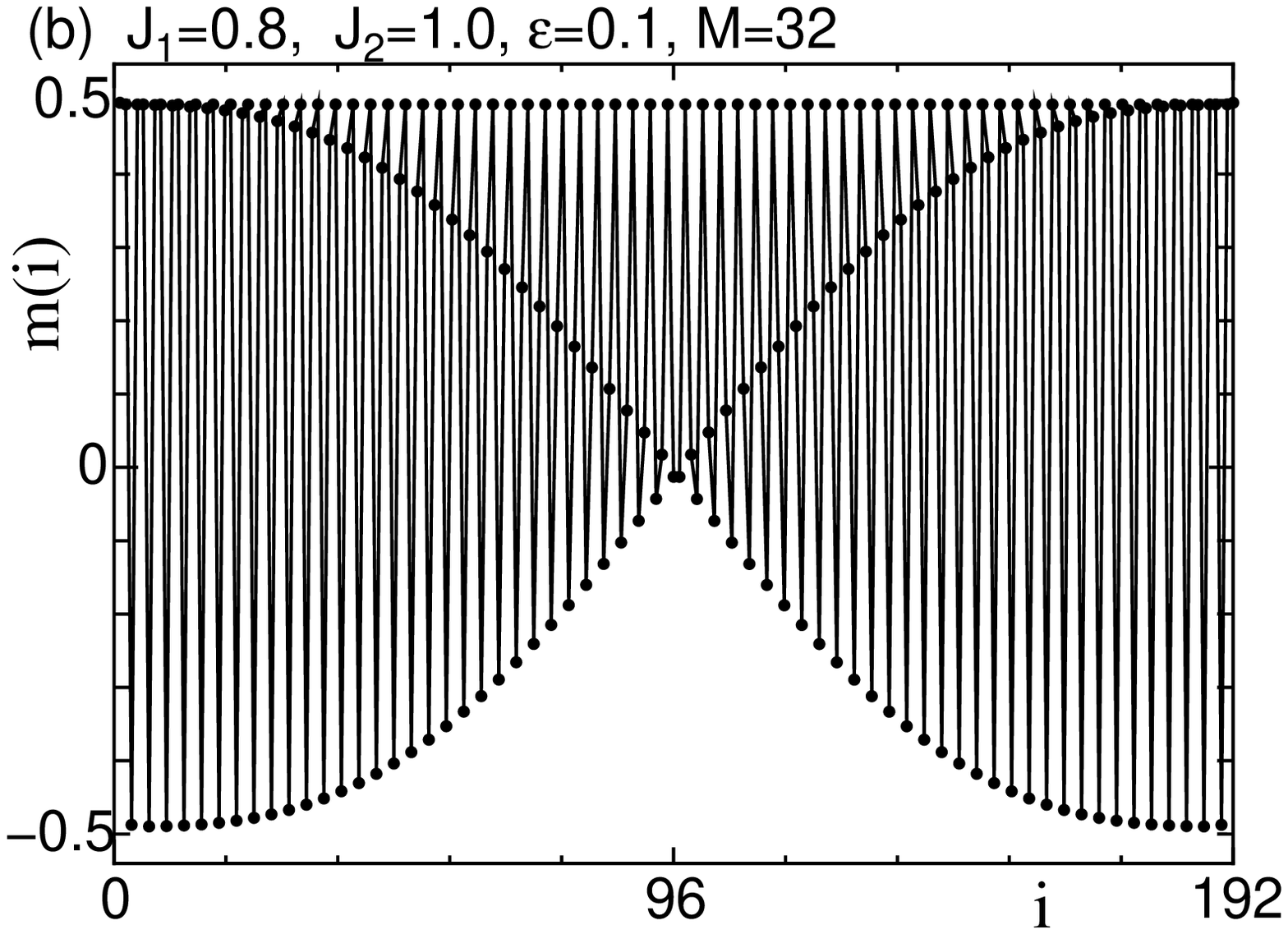,width=5.5cm}
\caption{The distribution of the local spin moment $m(i)$ at the 1/3 plateau state: (a) $J_1=1.0$ and $J_2=0.8$, and (b) $J_1=0.8$ and $J_2=1.0$. }
\label{pladist}
\end{figure}

According to the plateau quantization condition,  the period of the 1/3 plateau state must be $q=3\times$integer  with the spontaneous symmetry breaking of the translation.\cite{oshikawa}
The three kind of the N\'eel states clearly satisfy this condition.
However for the  OBC system, we have discussed that the $S^z=-2/3$ DW can be inserted in the 1/3 plateau state in the Ising limit.
In order to investigate the effect of the XY term on the defect, we calculate the distribution of the local moment along the chain with the DMRG method, which is shown in FIG.\ref{pladist}.
We can clearly see that the defect is inserted at the center of the chain(FIG.\ref{pladist}-(b)).
A precise computation yields that the defect  appears for $\alpha>0.923$, which is consistent with the result for the Ising limit.
Moreover, we note that the DW inserted plateau state survives  even for the isotropic zigzag  chain, for which the appearing point of the DW-inserted plateau state is  $\alpha \simeq 0.85$.

Here it should be remarked that the translational symmetry of period 3 for the 1/3 plateau is broken by the  inserted DW defect.  
However, this does not mean the break down of the plateau quantization condition that is valid for the infinite length chain.
The present DW  defect in the plateau is the single impurity problem in a finite size system, which can be neglected in the bulk limit.
Therefore, the DW defect in the plateau state does not conflict with the quantization theorem which is relevant for the infinite length chain.

\section{summary and discussion}

In this paper, we have studied the excitation around the 1/3 plateau state of the $S=1/2$ Ising-like zigzag XXZ chain.
We have explicitly constructed the low-energy excitations as the DWs separating the three-fold degenerating N\'eel-ordered spin arrays of the 1/3 plateau.
The important point is that the DWs are regarded as the quasi-particles carrying the fractional value of $S^z$:  the $S^z=\pm 1/3$ DW and its bound state having $S^z=\pm 2/3$.
Further, we have taken account of the quantum effect of  the XY term with the degenerating perturbation theory.
Then we have found that the low-energy dynamics of the DWs is significantly affected by the interplay between the single and double chain natures of the system originating from the zigzag structure of the model; 
The $S^z=\pm 1/3$ DW moves by using the NN interaction term, while $S^z=\pm 2/3$ DW bound state uses the NNN one.
Analyzing the formation mechanism of the DW bound state precisely, we have shown that the crossover of the dispersion curves in the low-energy region occurs at $\alpha=\alpha^*_+$ for the $S^z=1/3$ DW, and at $\alpha=\alpha^*_-$ for the $S^z=-1/3$ DW.
Moreover, this crossover of the dispersion curves certainly induces the exotic behavior of the magnetization curve:
The cusp singularity and the even-odd oscillation of the magnetization curve are switched with each other at $\alpha=\alpha^*_\pm$,  depending on the relative position of the dispersion curves of the free  $S^z=\pm 1/3$ DW and the $S^z=\pm 2/3$ DW bound state.
In particular, the fact that the usual $S^z=\pm 1$ excitation is represented as a combination of the $S^z=\pm 1/3 $ DW particles is  essential for the even-odd oscillation of the curve.
In fact the DW excitation combined with the hard-core bosonic particle picture has explained well the DMRG calculated magnetization curves.
Although the present argument is based on the degenerating perturbation theory of the first order with respect to $\varepsilon$,  the intrinsic properties of the DW excitations are vividly illustrated within the present theory, which is clearly supported by the DMRG results for the magnetization curve.

Here it is worthwhile to remark about the connection of the present DW picture to other view points for the magnetization curve.
As was seen above, the fractional $S^z$ DW is valid for the magnetization curve near the 1/3 plateau. 
On the other hand, we know that a good picture to describe the magnetization curve in high-field region is the spin wave excitation from the saturation field, which also explains the cusp singularity successfully\cite{cusp}.
Then both of the DW picture around the 1/3 plateau and the spin wave picture from the saturated state seem to capture the qualitative feature of the magnetization curve in the crossover region of these two pictures.
A similar situation is also considered for the low-field branch.
For $\varepsilon=0.1$ the ground state at the zero magnetic field is the N\'eel or double-N\'eel order. 
More interestingly, the dimer-singlet order is developed particularly around $\alpha=1/2$ as the system approaches isotropic point($\varepsilon\to 1$).
For the isotropic system,  the excitation at the zero field is represented as the DW separating the two-fold degenerating dimer-single states, for which the bound state of such DW is also reported\cite{zigzagss,brehmer}.
However the relationship between the DW picture at the 1/3 plateau and the DW based on other pictures has not been clarified yet.
The unified view for these complemental pictures may be a key problem to reveal the interaction effect between the DW particles in the zigzag-like  quantum spin chain.

In addition,  we would like to comment on the recently synthesized zigzag compound (N$_2$H$_5$)CuCl$_3$, which is considered as a isotropic zigzag chain of the coupling constant $J_1/J_2\simeq 0.25$.\cite{haginaru}
In this parameter region, the magnetization curve  does not have the 1/3 plateau.
However, the compound belongs to the region of the even-odd oscillation.
Of course the bulk magnetization curve does not show such oscillating behavior, but we can expect the dual structure of the DW and its bound state in a magnetic field.
According to the present theory, the magnetic excitation accompanying the $\pm 1$ change of the magnetization should be described by the combination of $S^z=\pm 1/3$ DW and $S^z=\pm 2/3$ DW.
Recalling that ESR or NMR measurements is theoretically described by $\langle S^+ S^-\rangle $ type correlation function, we can expect that the spectra of these experiments may resolve the DW excitation and its bound state.

Finally we wish to stress that our approach is quite general and is applicable to the plateau states represented by the Ising bases accompanying the spontaneously symmetry breaking of the translational invariance.
The DW particles having the fractional $S^z$ is essentially originating from the discreteness of the Ising bases and the degenerating plateau state;
If there is a $P$-fold degenerating plateau state, there is a possibility of a fractional DW excitation of $S^z=1/P$.
It is actually known that the $S=1$ Ising-like zigzag XXZ chain exhibits various long-period state in a magnetic field.\cite{kabu}
Our approach is useful for analysis of such a problem.
In addition, recently, a ${\cal N}$=2 super-symmetric 1D lattice model is proposed, where the very similar ordered state and the excitation having the fractional U(1)-charge are explicitly constructed.\cite{fendley}
It may be an interesting problem whether such mathematical model can be associated with the our explicit construction of the fractional $S^z$ excitation.


\acknowledgments

We would like to thank  T. Hikihara and M. Kaburagi for valuable comments.
K.O also thank  N. Maeshima, A. Koga and A. Kawaguchi for fruitful discussions.
This work is partially supported by a Grant-in-Aid for Scientific Research on Priority Areas (B) from the Ministry of Education, Culture, Sports, Science and Technology of Japan.



\begin{thebibliography}{99}
\bibitem{vaccumene} The energy of the DW particles is measured from that of the 1/3 plateau state of the PBC system.
\bibitem{position} In Sec. III, the DWs are labeled by $l$ or $m$ indicating the distance between the two DWs in the unit of three spins. 
However we here label the position of the DW as the site index in order to calculate the dispersion curve along the chain.
\bibitem{second} The second order perturbation leads the $S^z=1/3$ DW dispersion curve as $
\omega(k)=\frac{1}{3}(J_1+J_2)
+ \frac{2\varepsilon^2}{3} \left (\frac{ J_1^2}{J_2}+ \frac{J_2^2}{J_2}  \right)-\frac{\varepsilon^2}{2}J_1
+\frac{(2-\varepsilon)\varepsilon  J_1}{2} \cos 3k
-\frac{\varepsilon^2}{2}\frac{J_1^2}{J_2} \cos 6k
$.
However,  the qualitative feature of the dispersion curve near the bottom($k=\pi/3$) is not changed by the 2nd order perturbation, since the coefficient of $\cos 6k$ term is negative.
\bibitem{evensite} We have assumed $N$=even, so that the double N\'eel order($\uparrow\uparrow\downarrow\downarrow$) is realized at zero magnetic field with no defect.
If we use $N$=odd, the roles of the $M=$even step and $M=$odd step are exchanged.
\bibitem{fsolv} The value $\alpha=0.579$ is obtained by $\frac{1+\varepsilon}{2-\varepsilon}$ with $\varepsilon=0.1$, at which the dispersion curves extended from $u=\pi/6$ and $\pi/2$ connect with each other. We also determine $\alpha=0.605$ by solving $\frac{d^2}{d u^2} E^{(2)}(u)=0$ numerically.
\end{thebibliography}
\end{document}